\documentclass{article}
\usepackage{hyperref}
\usepackage{amsmath}
\usepackage{graphicx}
\usepackage{tikz}
\usepackage{float}
\usetikzlibrary{positioning, calc, decorations.pathreplacing}
\usepackage[utf8]{inputenc}
\usepackage[top=1in,bottom =1in, left=1in,right=1in]{geometry}
\usepackage{amsmath,amssymb,mathtools,bm,isomath,amsthm}
\usepackage{graphicx}
\usepackage{epstopdf}
\usepackage{color}
\usepackage{soul}
\usepackage{enumerate}
\usepackage{nicefrac}
\usepackage{enumitem}
\usepackage{mathrsfs}
\usepackage{scalerel,stackengine}
\usepackage{afterpage}
\usepackage{graphicx}
\usepackage{url}
\usepackage[affil-it]{authblk} 
\usepackage[toc,page]{appendix}
\usepackage{hyperref}
\usepackage{amsmath}
\usepackage{tikz}
\usepackage{float}
\usepackage{optidef}
\usepackage{algpseudocode}
\usetikzlibrary{positioning}
\usetikzlibrary{positioning, calc, decorations.pathreplacing}
\usetikzlibrary{matrix, positioning}
\usetikzlibrary{fit,positioning}
\usepackage[export]{adjustbox}
\usepackage{pgfplots}
\usepackage{graphicx}
\usepackage{epstopdf}  
\usepackage{flafter,placeins}
\usepackage{subcaption} 
\usepackage{listofitems}
\usepackage{tikz}
\usepackage{siunitx} 
\usepackage{amssymb} 
\usepackage[bottom]{footmisc} 

\usepackage{amsmath,color,soulutf8,longtable,colortbl,setspace,xspace,url,pdflscape,cite}

\usepackage[nolist]{acronym}

\usepackage{amsmath,color,soul,longtable,colortbl,setspace,xspace,url,pdflscape,cite}

\usepackage[nolist]{acronym}
\usepackage{fancyhdr}
\fancypagestyle{pagenumber}{\fancyhf{}\fancyhead[R]{\thepage}}

\makeatletter
\let\ps@plain=\ps@pagenumber
\makeatother

\usepackage{hyperref}
\makeatletter
\providecommand*{\toclevel@compteur}{0}
\makeatother
\usepackage{caption}
\usepackage{subcaption}
\usepackage{siunitx}
\usepackage{algorithm}  
\usepackage{amssymb} 
\usepackage[bottom]{footmisc} 

\newcommand{\x}{\boldsymbol{x}}

\pgfplotsset{compat=1.18} 
\begin{document}
	\setlength{\baselineskip}{15pt}
	
\title{Approximating electromagnetic fields in discontinuous media using a single physics-informed neural network}
\author[1]{Michel Nohra}
	\author[1]{Steven Dufour}
	\affil[1]{D\'epartement de math\'ematiques et de g\'enie industriel, Polytechnique Montr\'eal, Montr\'eal, Qu\'ebec, Canada, H3T 1J4}
	
	\date{\today}
	
	\maketitle
\begin{abstract}

\noindent 
{\bfseries Keywords:} PINN, Electromagnetism, Maxwell, Parametric PINN
  \end{abstract}
  Physics-Informed Neural Networks (PINNs) are a new family of numerical methods, based on deep learning, for modeling boundary value problems. They offer an advantage over traditional numerical methods for high-dimensional, parametric, and data-driven problems. However, they perform poorly on problems where the solution exhibits high frequencies, such as discontinuities or sharp gradients. In this work, we develop a PINN-based solver for modeling three-dimensional, transient and static, parametric electromagnetic problems in discontinuous media. We use the first-order Maxwell's equations to train the neural network.
We use a level-set function to represent the interface with a continuous function, and to enrich the network's inputs with high-frequencies and interface information. Finally, we validate the proposed methodology on multiple 3D, parametric, static, and transient problems.
\section{Introduction}

The study of electromagnetism in discontinuous media is an important problem found in numerous engineering applications, from antenna systems to nanoscale devices. For the majority of electromagnetic problems, analytical solutions are hard to obtain, and numerical methods, such as the finite element method, are used to approximate solutions. Due to advancements in computer hardware, neural networks have recently gained large popularity and have achieved remarkable results in numerous fields, from computer vision, where they have been used to detect cancer~\cite{nasser2019lung}, to language processing~\cite{goldberg2016primer}. 
Neural networks are increasingly used in scientific computing, for tasks like developing turbulence models from data \cite{turb1, turb2, turb3, turb4}, and simulating heat transfer problems \cite{HT1, ht2}. Notably, Physics-Informed Neural Networks (PINNs) are gaining traction among computational scientists. Introduced by Raissi et al. \cite{raissi}, PINNs integrate partial differential equations (PDEs) and all related information on their solutions into the neural network training. 
In electromagnetism, PINNs have been used to discretize both forward and inverse electromagnetic problems \cite{max2,max_net,meta_des,max2sG}. Kovacs et al~\cite{max2} used a PINN for the 2D magnetostatic forward problem of predicting the magnetic vector potential given a magnetization, and the inverse problem of identifying the magnetization from a given magnetic vector potential. The authors used two neural networks: the first predicts the magnetization distribution, while the second calculates the vector potential based on the magnetization obtained from the first network's output.
Another study done by Lim et al.~\cite{max_net} focuses on approximating the electric field distribution, given an electromagnetic lens shape using PINNs. The authors use the 2D $E$-formulation in the frequency domain to train the network. Their work also tackles the inverse problem of determining the lens shape responsible for an electric field distribution. The lens shape is parameterized by a decoder network, which is used to determine the lens shape that produces a given electric field distribution.
Similarly, Fang et al~\cite{meta_des} used two neural networks to estimate the electric field distribution from the material's permeability and permittivity, and vice versa.
Dong et al.~\cite{max2sG} worked on the $2$D magnetostatic problem with discontinuous media. The boundary conditions were weakly imposed, and Maxwell's equations formulation used to train the neural network was given by: 
\begin{align*}
 &\nabla \times \boldsymbol{H} = \boldsymbol{J}; \\
 &\mu \boldsymbol{H} = \nabla \times \boldsymbol{A}. 
\end{align*}
Using this formulation, the authors avoided the need to differentiate the discontinuous permeability, at the expense of adding a variable $\boldsymbol{A}$.

For electromagnetic problems, the presence of an interface introduces a discontinuity in the electromagnetic fields. Even though neural networks are universal approximators~\cite{uni_app,uni_approx}, a neural network will face difficulties and slow down convergence when approximating a high-frequency function, such as a discontinuous function, due to the spectral bias of neural networks~\cite{spec_bias,spec_bias2,NTK}.
Several remedies for the spectral bias have been proposed, such as the Fourier Feature Neural Networks (FFNN)~\cite{fourier_net,fourierpinn}. With the FFNN approach, the input vector $\x$ of the neural network is mapped to a high-dimensional vector, with richer frequencies, denoted as $\x_{\rm f}$, such that $$\x_{\rm f} = [\cos{(\x B )} \ , \  \sin{ (\x B)} ], $$
where $B$ is a uniformly distributed matrix with a user-defined range. The range of values in $B$ is important for effective convergence, and should ideally match the solution's frequency range. The Fourier mapping will not accelerate convergence if the frequency range is too low, and if the frequency range is too high, it will introduce high-frequency residuals that cause slow, or no convergence\cite{fourierpinn}.

Identifying the correct frequency range is particularly challenging for functions with varying frequencies across the computational domain. Such cases include solutions to boundary layer problems and interface problems, which are often smooth in some parts of the domain, but contain sharp gradients in other parts of the domain. For this reason, we propose the Interfacial PINN (I-PINN), a neural network suitable for problems with material interfaces.

With the I-PINN, the input vector of the neural network is $$\x_{\rm a}=[\x,\boldsymbol{f}(\x),\boldsymbol{n}(\x)], $$
where $\x$ contains the space-time coordinates, $\boldsymbol{f}(\x)=[f_1(\x),f_2(\x),f_3(\x),\ldots]$ is a set of functions that introduce high frequencies around the interface, and $\boldsymbol{n}(\x)$ is a vector field function that introduces topological information about the interface.
We use the level-set function to calculate $\boldsymbol{f}(\x)$ and $\boldsymbol{n}(\x)$, and a smooth approximation of the interface. We also strongly impose boundary and initial conditions using the method introduced in \cite{sukumar2022exact}.
In the remainder of this paper, we introduce Maxwell's equations and their various formulations, we then briefly introduce PINNs, we explain the proposed method, and we validate the proposed methodology using various three-dimensional parametric problems.

\section{Maxwell's equations}

Maxwell's equations are a set of four partial differential equations that describe the behavior of electromagnetic fields.
They describe how the electric and magnetic fields interact with each other and with matter.
For linear materials, the relations between the magnetic field $\boldsymbol{H}$, the magnetic flux $\boldsymbol{B}$, the electric field $\boldsymbol{E}$, and the electric displacement $\boldsymbol{D}$, are given by:
\begin{subequations}
\begin{align}
    \partial_t \mu\boldsymbol{H}+ \nabla \times\boldsymbol{E}= 0;\label{eqn:full_max_31} \\
    \partial_t \epsilon\boldsymbol{E}- \nabla \times\boldsymbol{H}+\boldsymbol{J}= 0; \label{eqn:full_max_32}\\
    \nabla \cdot \epsilon\boldsymbol{E}= \rho_c; \label{eqn:full_max_33}\\
    \nabla \cdot \mu\boldsymbol{H}= 0, \label{eqn:full_max_34}
\end{align}
\label{eqn:full_max_3}
\end{subequations}

with $\boldsymbol{B}=\mu \boldsymbol{H}$ and $\boldsymbol{D}=\epsilon \boldsymbol{E}$, where the permeability $\mu$ and permittivity $\epsilon$ are constants, $\rho_c$ is the charge density, and $\boldsymbol{J}$ is the current density, given by $\boldsymbol{J}= \sigma\boldsymbol{E}+ \boldsymbol{u} \times \mu \boldsymbol{H},$ where $\sigma $ is the conductivity, and $\boldsymbol{u}$ is the velocity of the electric charges. In this work, we will consider this velocity to be zero.

Various formulations exist for Maxwell's equations, under various assumptions and fields of interest. We start with the steady state assumption, where we consider that the dependent variables do not vary in time, i.e. $\frac{\partial \cdot}{\partial t} =0$. System~\eqref{eqn:full_max_3} then becomes:

\begin{subequations}
\begin{align}
    \nabla \times\boldsymbol{E}=0 ;\label{eqn:steady_max_T31} \\
    \nabla \times\boldsymbol{H}= \sigma\boldsymbol{E}; \label{eqn:steady_max_T32}\\
    \nabla \cdot \epsilon\boldsymbol{E}= \rho_c; \label{eqn:steady_max_T33}\\
    \nabla \cdot \mu\boldsymbol{H}= 0.\label{eqn:steady_max_T34}
\end{align}
\label{eqn:steady_max_T3}
\end{subequations}
Equations \eqref{eqn:steady_max_T3} can be formulated differently depending on the assumptions and the fields of interest, to list a few:
\begin{itemize}
    \item The electric potential field equation in conductive media,
    $$\nabla \cdot \sigma \nabla V =0,$$
    where $\boldsymbol{E}=\nabla V$;
    \item The electric potential field equation in non-conductive-media,
     $$\nabla \cdot \epsilon \nabla V =\rho_c,$$
     where $\boldsymbol{E}=\nabla V$;
     \item The magnetic scalar potential field equation in non-conductive media,
     $$\nabla \cdot \mu \nabla A =0 ,$$
     where $\boldsymbol{H}=\nabla A$;
     \item The magnetic vector potential field equation in conductive media,
     $$\nabla \times \frac{1}{\mu} \nabla \times \boldsymbol{A} = \boldsymbol{J}, $$
     where $\mu\boldsymbol{H}= \nabla \times \boldsymbol{A} $.     
\end{itemize}

For the low-frequency regime, where $\partial_t \epsilon\boldsymbol{E}<<\boldsymbol{J}$, the following quasi-static equations are obtained from equation \eqref{eqn:full_max_3}:
\begin{subequations}
\begin{align}
  \partial_t \mu\boldsymbol{H}+ \nabla \times\boldsymbol{E} = 0; \label{eqn:lowf_max_31}\\
   - \nabla \times\boldsymbol{H}+ \sigma\boldsymbol{E} =0; \label{eqn:lowf_max_32}\\
  \nabla \cdot \mu\boldsymbol{H}=0; \label{eqn:lowf_max_33}\\
  \nabla \cdot \epsilon\boldsymbol{E}= \rho_c, \label{eqn:lowf_max_34}
\end{align}
\label{eqn:lowf_max_3}
\end{subequations} 
  A common formulation of the quasi-static equations is obtained by replacing $\boldsymbol{E}$ from equation \eqref{eqn:lowf_max_32} in equation\eqref{eqn:lowf_max_31}, which gives
\begin{subequations}
\begin{align}
  \partial_t \mu\boldsymbol{H}+ \nabla \times \dfrac{1}{\sigma} \nabla \times\boldsymbol{H} &= \boldsymbol{F}_{\rm ext}; \\
  \nabla \cdot \mu\boldsymbol{H} &=0, 
\end{align}
\label{eqn:H-form_T3}
\end{subequations} 
referred to as the $H$-formulation. 
A similar formulation can be obtained for $\boldsymbol{E},$ 
$$ \mu \partial_t (\sigma \boldsymbol{E}) +  \nabla \times  \nabla \times\boldsymbol{E}=0.$$

\subsection{Interface conditions}
In the presence of a material interface, the permeability $\mu$, the permittivity $\epsilon$, and the conductivity $\sigma$ are discontinuous. This leads to the following interface conditions:
\begin{itemize}
    \item From the divergence-free property $\nabla \cdot \mu\boldsymbol{H}= 0 ,$  we obtain: $$\mu_1 \boldsymbol{H}_{1n} = \mu_2 \boldsymbol{H}_{2n},$$ where $\boldsymbol{H}_{1n}$ denotes the normal component of $\boldsymbol{H}$ on the first side of the interface, and $\boldsymbol{H}_{2n}$ is the normal component on the second side of the interface;
    \item By applying the divergence operator \eqref{eqn:lowf_max_32}, we obtain $$\nabla \cdot(- \nabla \times\boldsymbol{H}+ \sigma \boldsymbol{E})  =\nabla \cdot \sigma\boldsymbol{E}= 0 ,$$ leading to the interface condition $$\sigma_1 \boldsymbol{E}_{1n} = \sigma_2 \boldsymbol{E}_{2n};$$ 
    \item  From equation \eqref{eqn:lowf_max_32}, we obtain $\boldsymbol{H}_{1t} - \boldsymbol{H}_{2t} = \boldsymbol{K}_f,$ where $\boldsymbol{K}_f$ is the surface current density on the interface, and $\boldsymbol{H}_{1t}$ and $\boldsymbol{H}_{2t}$ are the tangential components of $\boldsymbol{H}$ on both sides of the interface;
    \item From equation~\eqref{eqn:lowf_max_31}, we obtain that $\boldsymbol{E}_{1t} = \boldsymbol{E}_{2t}.$
\end{itemize}

\section{Physics-informed neural networks}

In this section, we describe the components of a neural network, and how a neural network can be used to approximate a solution to a boundary value problem using the PINN introduced in~\cite{raissi}.
We start with a basic example to understand neural networks. Consider a network with just one layer. Here, $\x$ is the input and $\x_1$ is the output, with respective dimensions $n$ and $m$. The output is defined as
$$\x_1 = \psi (W \x + \boldsymbol{b}),$$
where $W$ is the weight matrix of dimensions $m \times n$, $\boldsymbol{b}$ is the bias vector of dimension $m$, and $\psi(\boldsymbol{z})$ is the activation function. The activation function is user-defined, and can take various forms, such as a cosine or a hyperbolic tangent function, and is applied element-wise on a vector $\boldsymbol{z}$.

We can stack multiple layers to form a multi-layer feedforward neural network, where the output from one layer is the input to the next layer. The overall output of a feedforward neural network becomes \[ N(\x) = \psi_\ell \big( W_\ell  \psi_{\ell-1}( W_{l-1} \psi_{\ell-2}(\cdots (\psi_1(W_1 \x + \boldsymbol{b}_1  )  )   \cdots) +\boldsymbol{b}_{\ell-1}  )       +\boldsymbol{b}_\ell     \big).  \]

\tikzstyle{mynode}=[thick,draw=blue,fill=blue!20,circle,minimum size=22]
\begin{figure}
    \centering
    \includegraphics{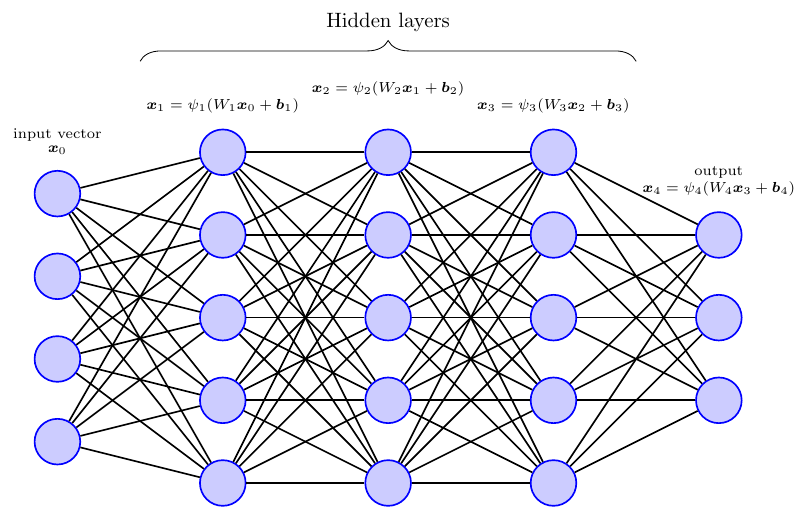}
    \caption{Feedforward neural network.}
    \label{fig:FFNN_T3}
\end{figure}
Building a more complex neural network is possible by changing the connections from layer to layer. 
After we build the neural network with the desired architecture and activation function, we train it to produce the desired output by tuning its parameters $\boldsymbol{\zeta}$, which is a vector containing all the weights and biases. Training is formulated as an optimization problem, where we minimize a user-defined loss function $L$ that measures the discrepancy between the network's output and the target output.
The optimization problem is formulated as
\begin{equation*}
    \min_{{\boldsymbol{\zeta}}}\ L(N_{\boldsymbol{\zeta}}(\x), D(\x)),
\end{equation*}
where $N_{\boldsymbol{\zeta}}(\x)$ denotes the neural network's output, $D(\x)$ is the desired output, and $L$ is the chosen loss function, such as the Mean Squared Error (MSE).

While various optimization methods could address this minimization problem, in the context of this work, we will focus only on gradient-based optimization methods. These algorithms exploit the property that the gradient of a function points in the direction of the function's steepest ascent. Therefore, to minimize the function, one would generally move in the direction opposite to the gradient. Popular gradient-based methods include Stochastic Gradient Descent (SGD), Limited-memory Broyden-Fletcher-Goldfarb-Shanno (L-BFGS)~\cite{bollapragada2018progressive}, and Adaptive Moment Estimation (Adam)~\cite{adam}. In neural networks, the gradient is calculated using the back-propagation algorithm, which use the chain rule to compute the partial derivatives of the network's output with respect to the weights, biases, and the network's input.
With the original PINN introduced by Raissi et al. \cite{raissi}, we use a feedforward neural network to approximate the solution to a boundary-value problem.
Consider a domain $\upOmega$ with the boundary $\mathrm{\Gamma} = \partial \upOmega$. We aim at approximating the solution $u(\x)$ of the boundary-value problem 
\begin{equation*}
    \begin{aligned}
         &R(\x,u ) =0\ , \ \forall \x \in \upOmega ;\\
         &B(\x ,u) =0\ , \ \forall \x \in \mathrm{\mathrm{\Gamma}},
    \end{aligned}
\end{equation*}
where $R$ is the residual of a PDE, and $B$ is the boundary condition we want to impose on $u(\x).$
In the vanilla PINN, we approximate the solution $u(\x)$ using a feed-forward neural network $N_{\boldsymbol{\zeta}}(\x)$, and we train the neural network parameters ${\boldsymbol{\zeta}}$ by minimizing the loss function defined as 
\begin{align*}
L &= \alpha_1 \sum_{i=1}^{n_\upOmega} R\ (\x_\upOmega^i, N_{\boldsymbol{\zeta}}(\x_\upOmega^i) \ )^2 + \alpha_2 \sum_{j=1}^{n_\mathrm{\Gamma}} B \ (\x_\mathrm{\Gamma}^j, N_{\boldsymbol{\zeta}}(\x_\mathrm{\Gamma}^j) \ )^2 + \alpha_3 \sum_{k=1}^{n_{\rm D}}\ ( N_{\boldsymbol{\zeta}}(\x_{\rm D}^k) - u(\x_{\rm D}^k) \ )^2  \\[2mm]
&= \alpha_1 \|  R\ (\x_\upOmega, N_{\boldsymbol{\zeta}}(\x_\upOmega) \ ) \| + \alpha_2 \|B \ (\x_\mathrm{\Gamma}, N_{\boldsymbol{\zeta}}(\x_\mathrm{\Gamma}) \ ) \| + \alpha_3 \|\ ( N_{\boldsymbol{\zeta}}(\x_{\rm D}) - u(\x_{\rm D}) \ ) \|,
\end{align*}
where $\{\x_\upOmega^i\}_{i=1}^{n_\upOmega}$ is a set of $n_\upOmega$ points that belong to $\upOmega$, $\{\x_\mathrm{\Gamma}^i\}_{i=1}^{n_\mathrm{\Gamma}}$ is a set of $n_\mathrm{\Gamma}$ points that belong to $\mathrm{\Gamma}$,  $\{\x_{\rm D}^i\}_{i=1}^{n_{\rm D}}$ is a set of $N_{\rm D}$ points where the solution is known (experimental measurements), $\alpha_i$ are weighting parameters that are user-defined, and $\|\cdot\|$ is the discrete $L2$ norm.

\section{Proposed methodology for the electromagnetic interface problem}
In this work, we use a PINN to approximate the solution to Maxwell's equations. 
In the original PINN introduced by Raissi et al. \cite{raissi}, the boundary conditions are weakly imposed. This weak imposition of boundary conditions can result in slow convergence during training, and could create local minima that do not satisfy the boundary conditions, nor the governing equations. The usual remedy is to add weight parameters for each term in the loss function, but the optimal value of these weights is not easy to determine.  
To avoid these problems, we will strongly impose boundary and initial conditions, using the neural network architecture proposed by Sukumar et al. \cite{sukumar2022exact}.

We will base our choice of Maxwell's equations formulation on the findings of Nohra et al. \cite{nohra2}, where we concluded that a first-order formulation offer better convergence properties and shorter computational time than a second-order formulation for interface problems. For this reason, we will train the PINNs using the first-order system of equations \eqref{eqn:steady_max_T3} for steady-state problems, and the system of equations \eqref{eqn:lowf_max_3} for transient problems. 
In numerous applications, the material interface is represented with a marker variable, such as for magnetohydrodynamic multi-fluid flows, where the use of the level-set method is common. For this reason, we will approximate the discontinuous physical quantities, such as the permeability and the conductivity, using a continuous approximation of the Heaviside function calculated using a level-set function $F$, which is a signed distance function to the interface. In this work, we use the sigmoid function to approximate the Heaviside function $\hat{H}=S(\alpha F)$, where $\alpha$ is a user-defined parameter that determines the sharpness of the approximation, and $S$ is the sigmoid function. With this approximation, the physical quantities will be expressed by: 
\begin{align*}
    & \mu = \mu_1 (1-\hat{H}) + \hat{H} \mu_2; \\
    & \sigma = \sigma_1 (1-\hat{H}) + \hat{H} \sigma_2.
\end{align*}
Using a smooth representation of the Heaviside function, the interface conditions will be automatically satisfied if the underlying PDEs are satisfied.
The loss function that we will use to train the PINN for the transient case is therefore given by
\[L= \|\partial_t \mu\boldsymbol{H}+ \nabla \times\boldsymbol{E}  \| + \|- \nabla \times\boldsymbol{H}+ \sigma\boldsymbol{E} \| + \| \nabla \cdot (\mu\boldsymbol{H}) \|  + \|\nabla \cdot (\epsilon\boldsymbol{E}) \|, \]
and for the steady-state case, we have
\[L= \| \nabla \times\boldsymbol{E}  \| + \|- \nabla \times\boldsymbol{H}+ \sigma\boldsymbol{E} \| + \| \nabla \cdot (\mu\boldsymbol{H}) \|  + \|\nabla \cdot (\epsilon\boldsymbol{E}) \|. \]

The neural network architecture described in \cite{nohra2} uses a form of overlapping domain decomposition. It uses a bump function to introduce the sharp gradients at the interface. The approximated vector field is then expressed as $\boldsymbol{v}=\boldsymbol{f}_1  +\hat{H}\boldsymbol{f}_2 $, where $\boldsymbol{f}_1$ and $\boldsymbol{f}_2$ are the outputs produced by the neural network. Although it showed promising results, and an improvement over the traditional PINN, it still faces several issues. The first issue is that it does not introduce additional information on the interface topology that helps the neural network converge faster during training. The second issue is that it does not scale well with multiple interfaces, because we need to increase the network size with the number of interfaces. The third issue is that it does not bypass the frequency bias of neural networks, making it increasingly difficult to model the interface with a sharper representation.

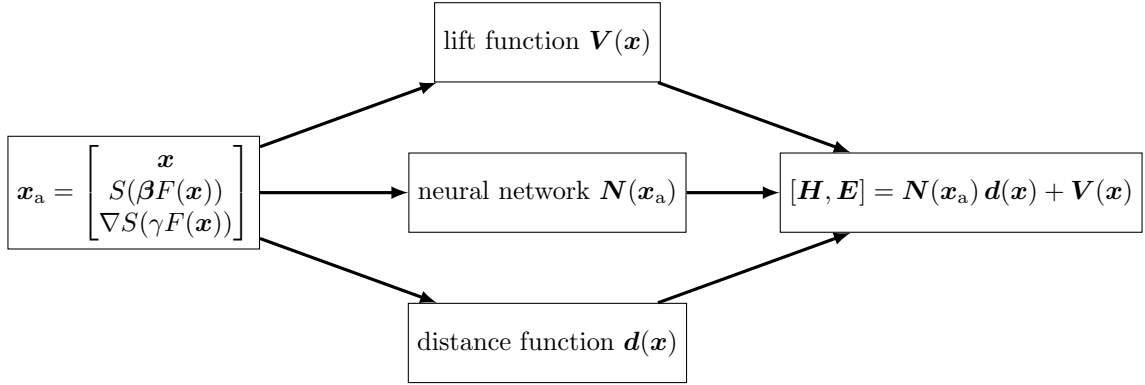
\begin{figure}[th]
\centering

\tikzstyle{block} = [draw, fill=white, rectangle, 
    minimum height=3em, minimum width=6em]
\tikzstyle{sum} = [draw, fill=white, circle, node distance=1cm]
\tikzstyle{input} = [coordinate]
\tikzstyle{output} = [coordinate]
\tikzstyle{pinstyle} = [pin edge={to-,thick,black}]

\begin{tikzpicture}[auto, node distance=2cm,>=latex]

    \node [block, name=input] (controller) {$\boldsymbol{x}_{\rm a} = \begin{bmatrix}
    \bm{x} \\
    S(\bm{\beta} F(\bm{x})) \\
    \nabla S(\gamma F(\bm{x}))
\end{bmatrix}
$};
    
    \node [block, right of=controller,
            node distance=5.5cm] (system) {neural network $\boldsymbol{N}(\boldsymbol{x}_{\rm a})$};
            
    \node [block, above of=system] (boundary) {lift function $\boldsymbol{V}(\boldsymbol{x})$};
    \node [block, below of=system] (distance) {distance function $\boldsymbol{d}(\boldsymbol{x})$};
    
    \node [block, right of=system,node distance=5.5cm] (sol) {$[\boldsymbol{H},\boldsymbol{E} ]=\boldsymbol{N}(\boldsymbol{x}_{\rm a})\, \boldsymbol{d}(\boldsymbol{x}) + \boldsymbol{V}(\boldsymbol{x})$};

    \draw [->,very thick] (controller) -- node[name=u] {} (boundary);
    \draw [->,very thick] (controller) -- node[name=u] {} (system);
    \draw [->,very thick] (controller) -- node[name=u] {} (distance);
    
    \draw [->,very thick] (system) -- node[name=u] {} (sol);
    \draw [->,very thick] (boundary) -- node[name=u] {} (sol);
    \draw [->,very thick] (distance) -- node[name=u] {} (sol);
  
\end{tikzpicture}

    \centering
    
    \caption{The architecture of I-PINN.}
    \label{fig:net_arch_M3}
\end{figure}

In this work, we will follow a different approach, where we introduce high frequencies near the interface, and include information on the interface topology in the PINN's input. 
We do so by adding $S(\boldsymbol{\beta} F)$ and $\nabla S(\gamma F)$ to the neural network's inputs, where  $\boldsymbol{\beta}$ and $\gamma$ are user-defined parameters.
This allows us to effectively achieve the following:
\begin{itemize}
    \item First, adding $ S(\boldsymbol{\beta} F)$ to the input introduces high frequencies only in the vicinity of the interface. Second, we are adding a marker for the PINN to distinguish the subdomains on each side of the interface;
    \item  $ \nabla S(\gamma F)$ represents a vector field normal to the interface. Adding the normal vectors introduces additional information about the interface shape, making it easier for the PINN to converge, and it makes it easier for the network to distinguish between the normal and tangential components of the output vectors.
\end{itemize}
The final architecture of the neural network is illustrated in figure~\ref{fig:net_arch_M3}.

\section{Numerical Results}
The proposed methodology is validated using a set of parametric problems, where the geometry and physical quantities are given in nondimensional units.

\subsection{Steady-state parametric problem of a sphere inside a unit cube}
\begin{figure}[H]
\begin{subfigure}[t]{0.5\textwidth}
\centering

    \includegraphics[width=\textwidth,height=\textwidth]{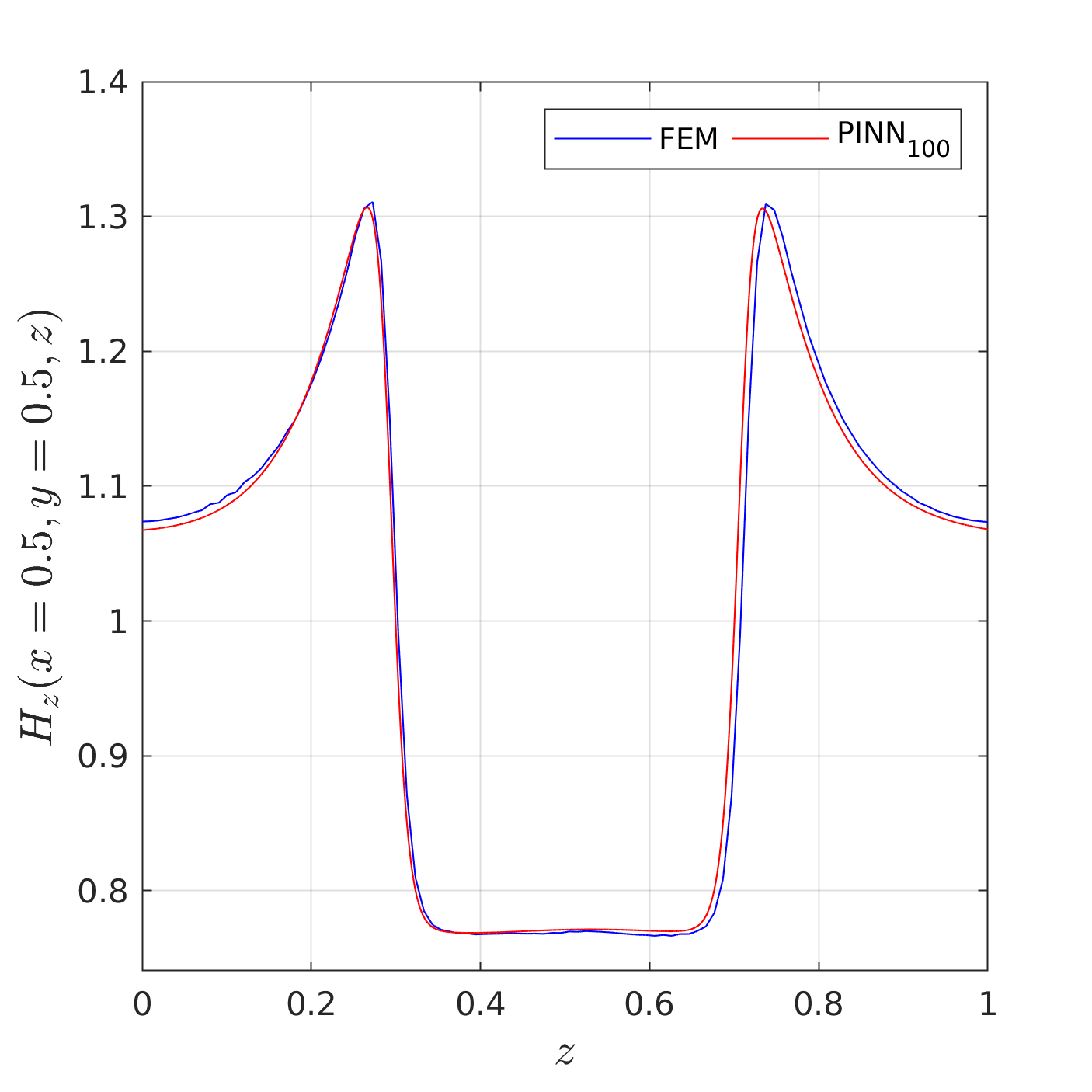}
\caption{}
\end{subfigure}
\begin{subfigure}[t]{0.5\textwidth}
\centering
    \includegraphics[width=\textwidth,height=\textwidth]{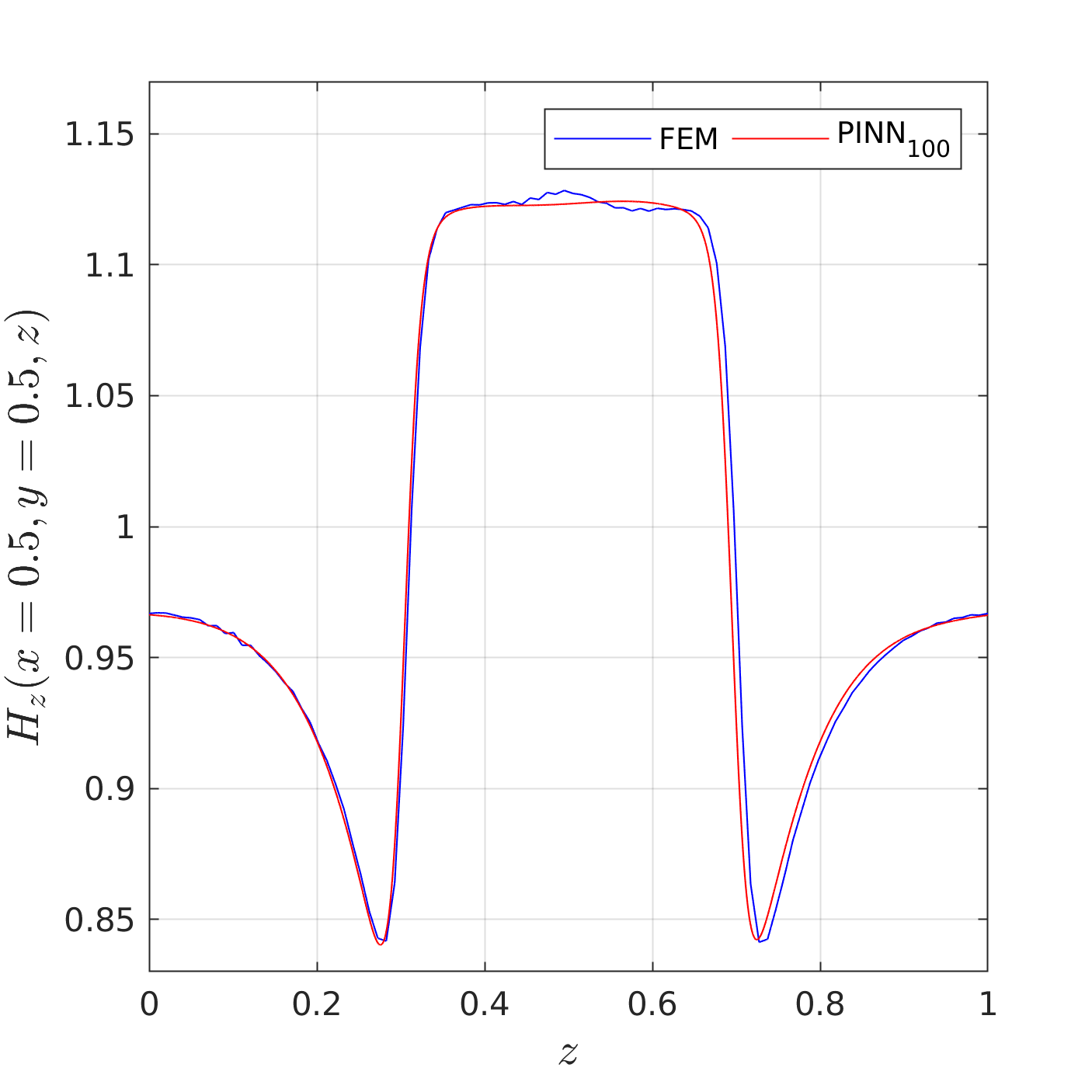}
\caption{}
\end{subfigure}

    \caption{The $z$-component of the magnetic field $H_z$ obtained using the proposed PINN method after 1000 Adams iterations and 2000 L-BFGS iterations, and $H_z$ obtained using the FEM, for the steady-state parametric problem of a sphere inside a unit cube: (a) $\mu_{\rm out} =0.5$, (b) $\mu_{\rm out} =1.5$.}
    \label{fig:A2_disc_mu}
  \end{figure}

We verify the developed methodology using the parametric steady-state magnetic problem consisting of a sphere with radius $r = 0.2$, located at the center of a unit cube. The Dirichlet boundary condition $\boldsymbol{H}=[0,0,1]$ is enforced on the four sides of the cube, and the Dirichlet boundary condition $\boldsymbol{E} = \boldsymbol{0}$ is enforced on the bottom and top sides of the cube. The permeability outside the sphere is set between $0.5 <\mu_{\rm out} < 1.5$, and the permeability inside the sphere is $\mu_{\rm in} = 1 $. The permeability
$\mu$ is approximated using $$\mu = S(100 F)  (\mu_{\rm out} -1) + 1,  $$
and the input to the PINN is $[\x,\mu_{\rm out},S(\boldsymbol{\beta} F), \nabla S(50F) ]$  with $\boldsymbol{\beta}= [50,100,150,200]$.
We compare, in figure~\ref{fig:A2_disc_mu}, the results obtained using the proposed PINN with the results obtained using the finite element method, for different values of $\mu_{\rm out}$.

\begin{figure}[H]
\begin{subfigure}[t]{0.5\textwidth}
\centering

    \includegraphics[width=\textwidth,height=\textwidth]{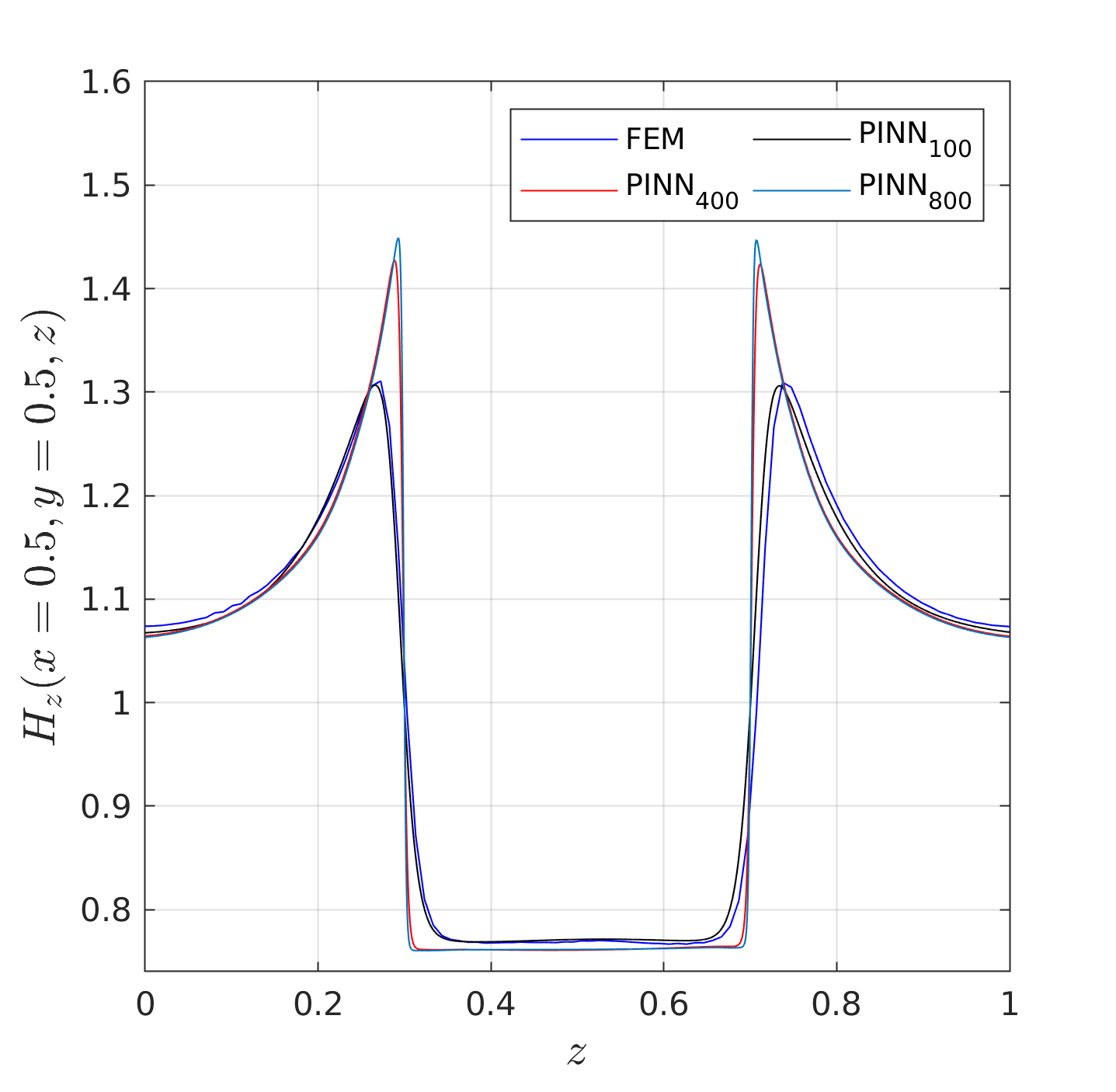}
\caption{}
\end{subfigure}
\begin{subfigure}[t]{0.5\textwidth}
\centering
    \includegraphics[width=\textwidth,height=\textwidth]{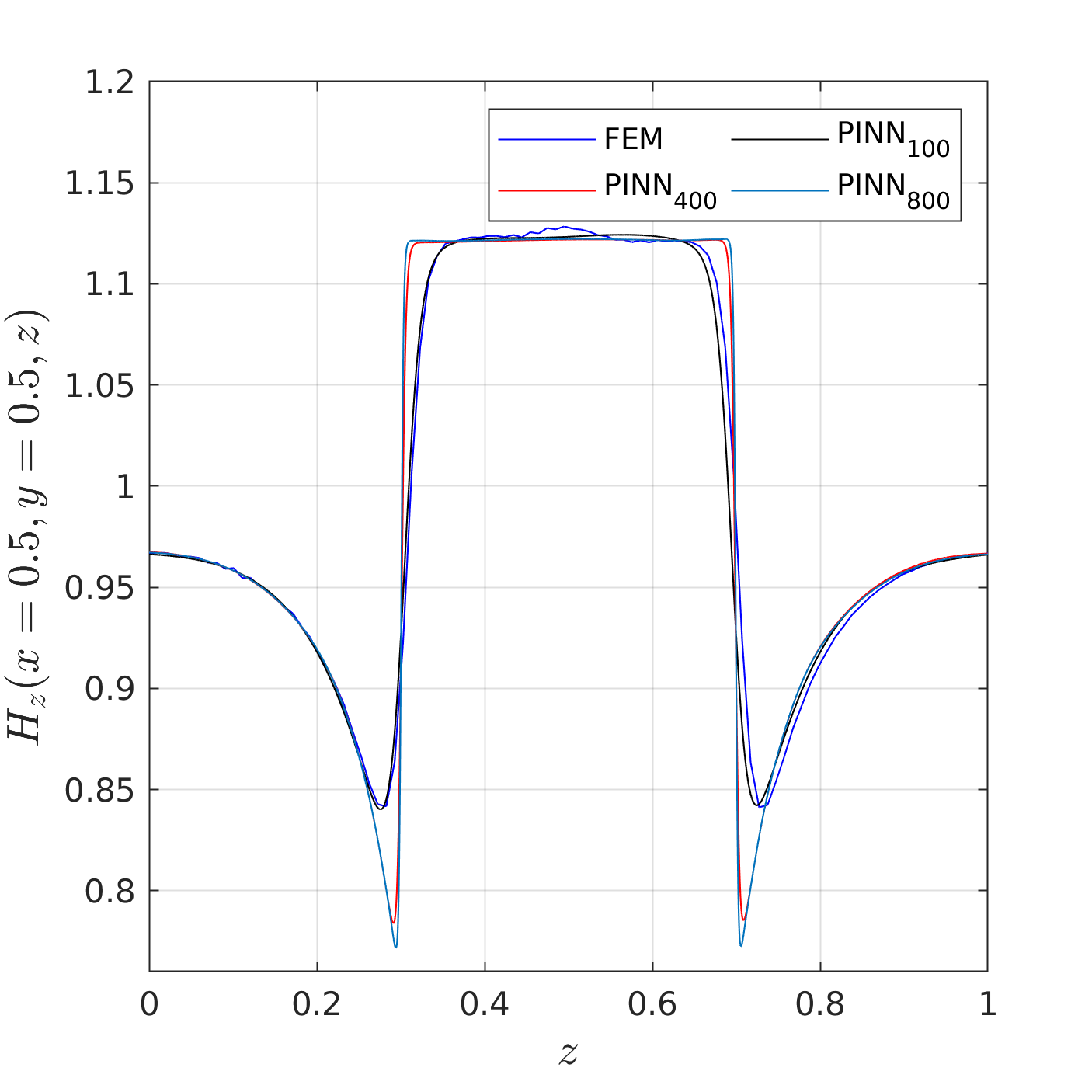}
\caption{}
\end{subfigure}

    \caption{The $z$-component of the magnetic field $H_z$ at $x=y=0.5$, using the approximate Heaviside function of $\hat{H}=S(\alpha F)$, with $\alpha=100, 400$ and $800$, for the parametric steady-state problem of a sphere inside a unit cube: (a) $\mu_{\rm out}=0.5$ (b) $\mu_{\rm out} = 1.5$. } 
    \label{fig:2int_disc_mu}
  \end{figure}

\begin{figure}[ht]
  \centering
    \includegraphics[height=7cm]{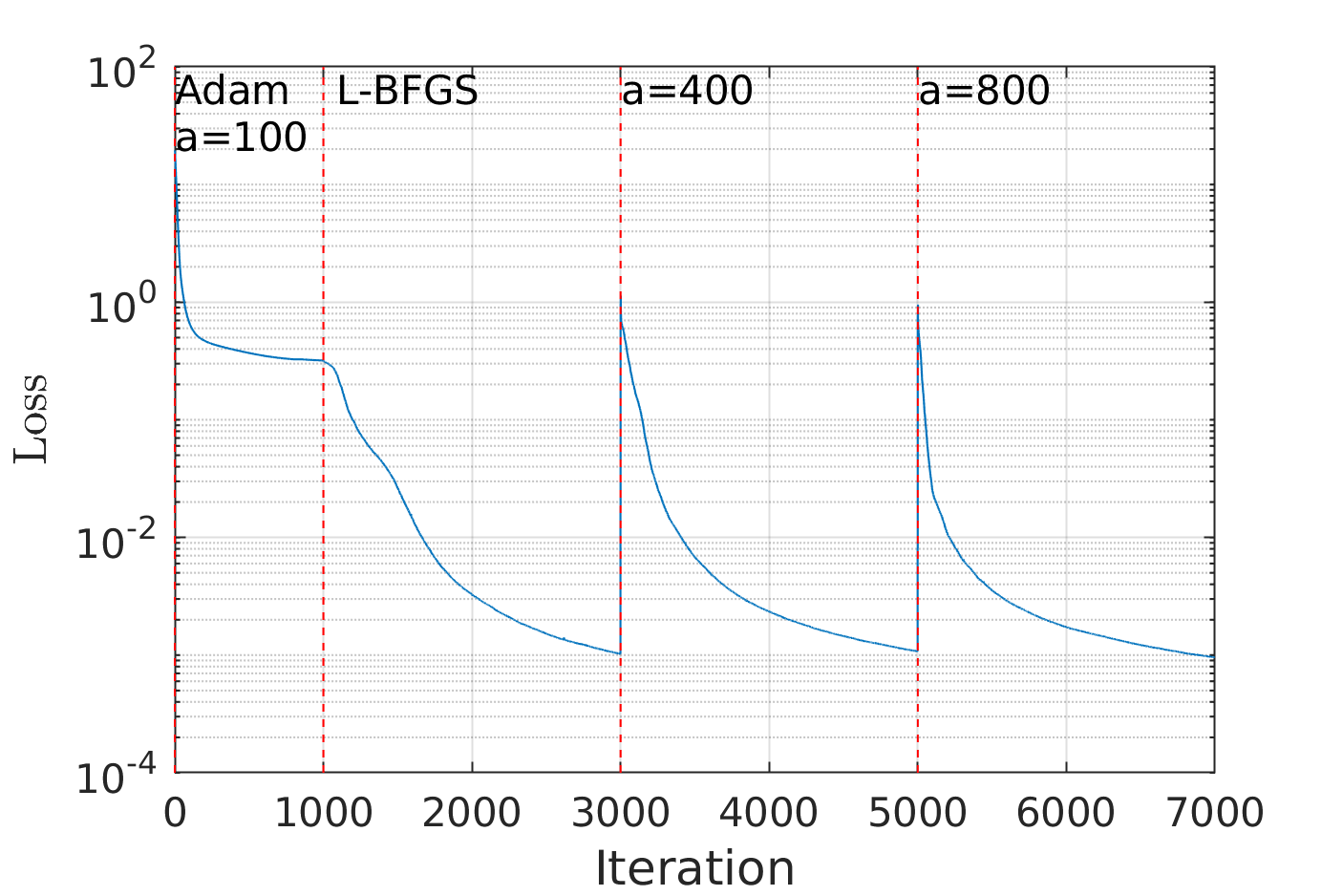}
    \caption{The loss function during training, for the parametric steady-state magnetic problem of a sphere inside a unit cube with curriculum learning, where the Heaviside function is approximated using $a=100$, and the network is trained with 1000  Adams iterations and 2000 L-BFGS iterations, then for another 2000 L-BFGS iterations with $a=400$, and another 2000 L-BFGS iterations for $a=800$.}
    \label{fig:A2_loss}
  \end{figure}

Next, we retrain the same neural network with a steeper approximation of the Heaviside function $\hat{H} = S(a F) $, first with $a=400$, and then with $a=800$. This training method is known as curriculum training, where the PINN is trained for a simple problem and then successively retrained for more complex problems. 
We compare the results obtained with the proposed PINN for different values of $a$, with the results obtained with the FEM in figure \ref{fig:2int_disc_mu}, and in figure \ref{fig:A2_loss} we show the loss function during the training stages.


The results of figure~\ref{fig:2int_disc_mu} show the effectiveness of the proposed neural network in capturing sharp gradients, and the practicality of the method, since only 3000 iterations were needed to obtain the first set of results for $a=100,$ and 7000 total iterations to obtain an approximation with a sharp interface representation.

\subsubsection{Problem of a sphere inside a unit cube using Fourier Feature PINN}

\begin{figure}[ht]
\begin{subfigure}[t]{0.5\textwidth}
\centering    \includegraphics[width=\textwidth,height=\textwidth]{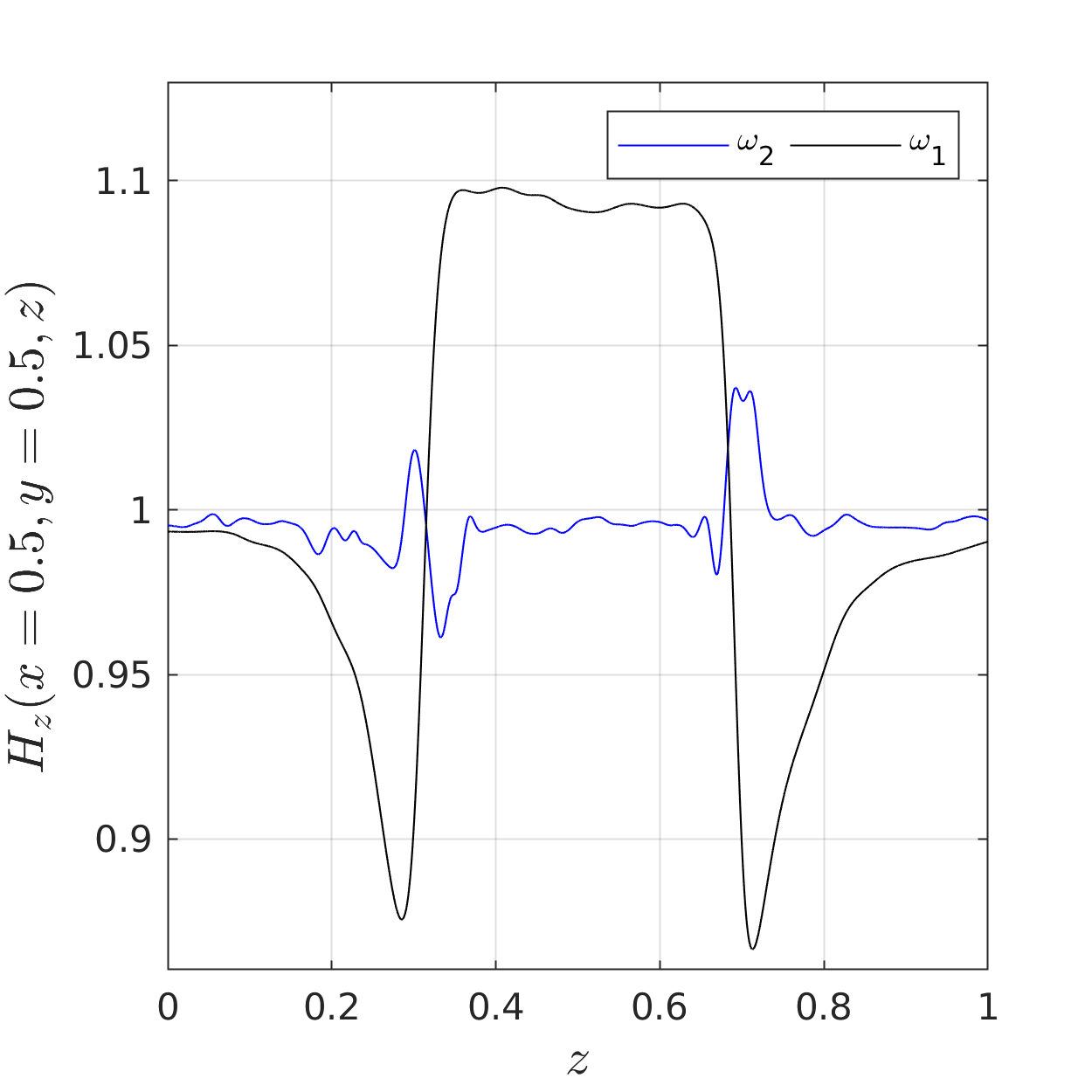}
\caption{}
\end{subfigure}
\begin{subfigure}[t]{0.5\textwidth}
\centering
    \includegraphics[width=\textwidth,height=\textwidth]{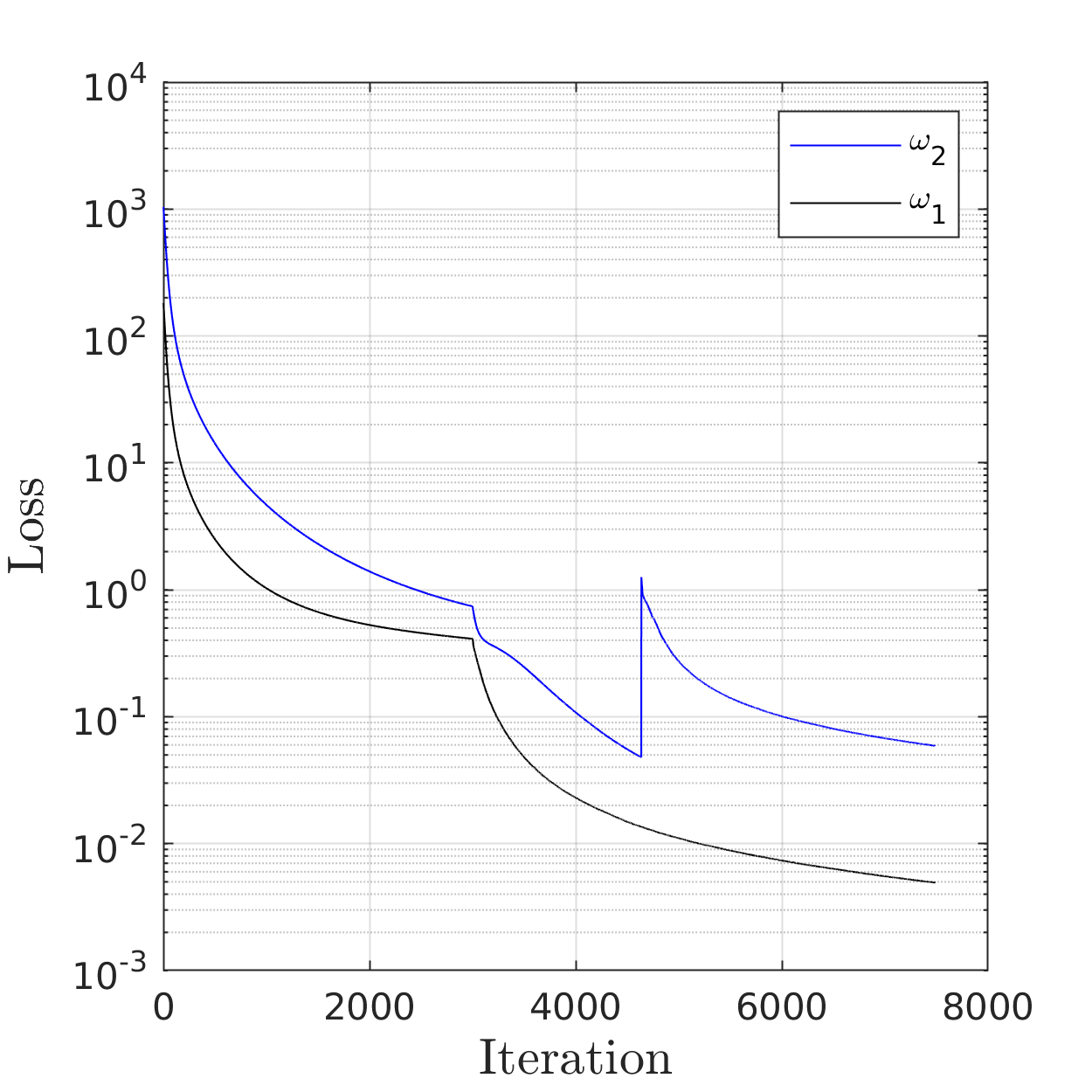}
\caption{}
\end{subfigure}

    \caption{The results obtained using the FF-PINN with a frequency range of $0<\omega_1<5\pi$ and $0<\omega_2<10\pi$, for the steady-state parametric problem of a sphere inside a unit cube: (a) $H_z$ at $x=y=0.5$, for $\mu_{\rm out}=1.5$; (b) the loss function during training. }
    \label{fig:FF-PINN}
  \end{figure}

As a demonstration of the importance of introducing high frequencies locally, we approximate the solution to the problem proposed above using the Fourier Feature PINN (FF-PINN). 
The results obtained using two FF-PINN are illustrated in figure \ref{fig:FF-PINN}. The first FF-PINN has a frequency range of $0<\omega_1<5\pi$, and the second FF-PINN has a frequency range of $0<\omega_2<10\pi$. The Heaviside function is approximated using $a=100$. 
The networks are trained with 3000 Adams iterations and 4500 L-BFGS iterations.
As we can see in figure \ref{fig:FF-PINN}, introducing high frequencies on the entire domain will slow down the convergence, and introduce high-frequency residuals in the smooth part of the solution, leading to erroneous results.

\subsubsection{Problem of a sphere inside a unit cube using the first-order and the second-order formulations}

\begin{figure}[H]
\begin{subfigure}[t]{0.5\textwidth}
\centering

    \includegraphics[width=\textwidth,height=\textwidth]{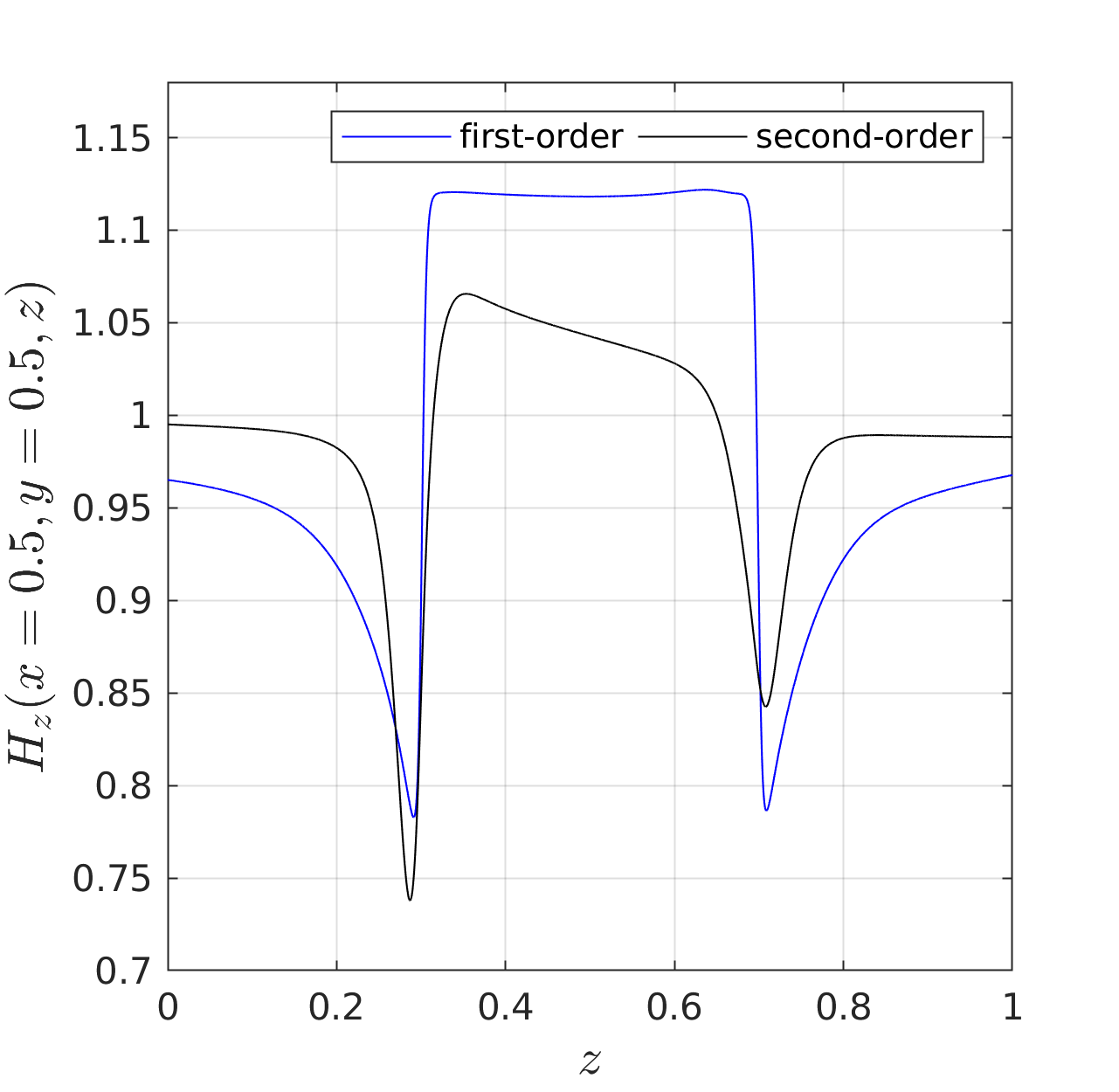}
\caption{}
\end{subfigure}
\begin{subfigure}[t]{0.5\textwidth}
\centering
    \includegraphics[width=\textwidth,height=\textwidth]{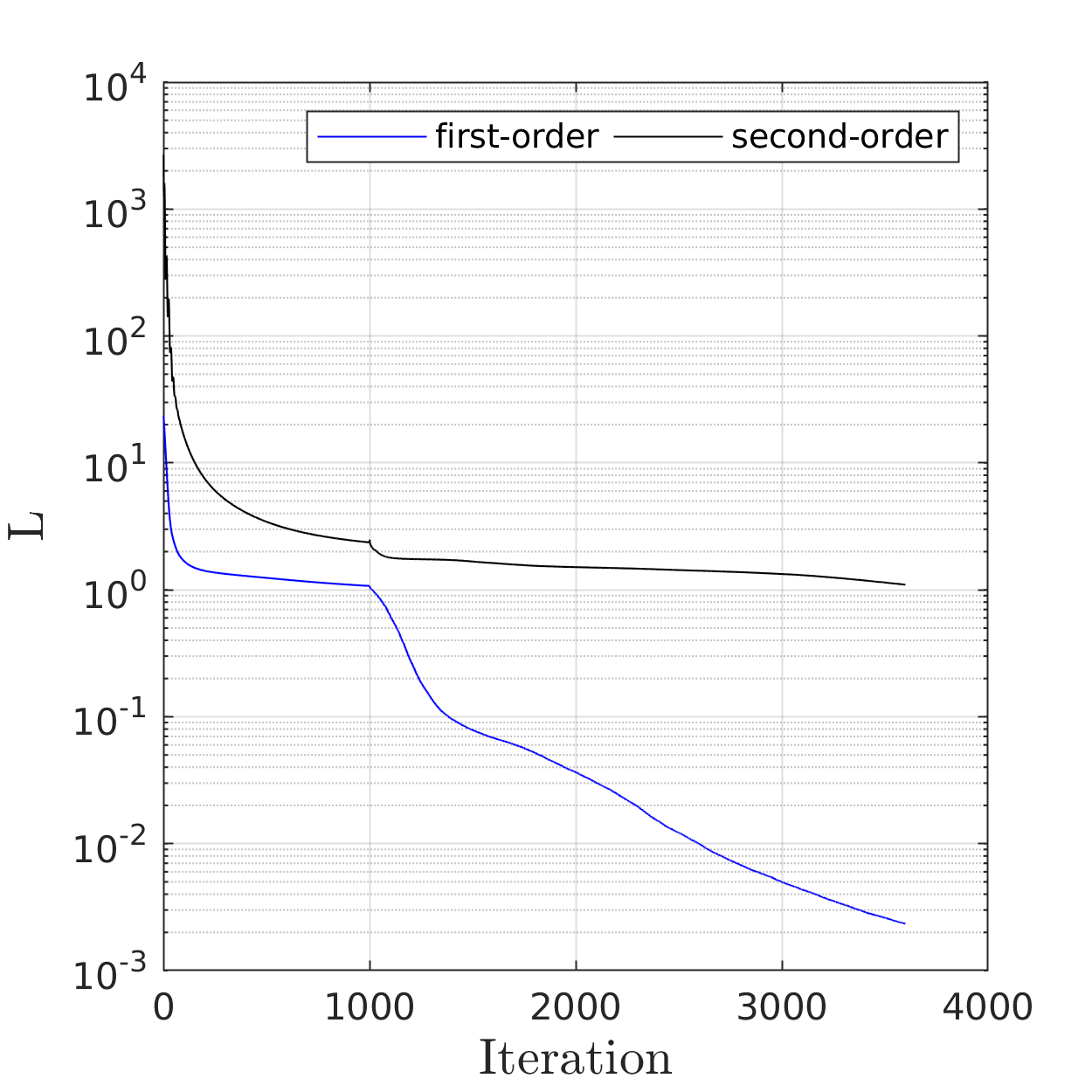}
\caption{}
\end{subfigure}

    \caption{The results obtained for the steady-state parametric problem of a sphere inside a unit cube using the first-order and second-order formulation: (a) $H_z$ at $x=y=0.5$, for $\mu_{\rm out}=1.5$; (b) the loss function during training.}
    \label{fig:order-PINN}
  \end{figure}
  
To show the effectiveness of using the first-order formulation, as opposed to the second-order formulation, we train the proposed neural network architecture with an approximate Heaviside function $\hat{H}=S(400F)$, on the parametric steady-state problem of a sphere inside a unit cube. The PINNs are trained using the first-order system of equations \eqref{eqn:steady_max_T3} and the second-order formulation \eqref{eqn:H-form_T3}, for the same number of iterations. We use the Adam optimizer for 1000 iterations, and the L-BFGS optimizer for 2600 iterations. As we can see in figures~\ref{fig:order-PINN}, the PINN trained with the second-order formulation hits a plateau and has difficulty converging, contrary to the PINN trained with the first-order formulation. It is worth noting that the cost of training using the second-order formulation is almost doubled when compared to the first-order formulation.

\subsubsection{Problem of a sphere inside a uniform magnetic field}

In the next experiment, we place a sphere of radius $r=0.2$ inside a cube with sides $\ell=2$. 
We impose a Dirichlet boundary condition on the magnetic field $\boldsymbol{H}=[0,0,1]$ on the four sides of the cube, and we impose a zero Dirichlet boundary condition on the electric field $\boldsymbol{E} = \boldsymbol{0}$ on the bottom and top sides of the cube.
Since the diameter of the sphere is small compared to the computational domain, the problem becomes similar to a metal sphere inside a uniform magnetic field. If the outside magnetic field is $\boldsymbol{H} = [0,0,H_z^0]$, the exact solution for $\boldsymbol{H}$ outside the sphere is then given by 
\[
  \boldsymbol{H} =  [
H_z^0 \dfrac{\mu_{\rm in}-\mu_{\rm out} }{\mu_{\rm in}+2\mu_{\rm out} } \dfrac{R^3}{r^5}3xz \ , \ 
H_z^0 \dfrac{\mu_{\rm in}-\mu_{\rm out} }{\mu_{\rm in}+2\mu_{\rm out} } \dfrac{R^3}{r^5}3yz \ , \ 
H_z^0 + H_z^0 \dfrac{\mu_{\rm in}-\mu_{\rm out} }{\mu_{\rm in}+2\mu_{\rm out} } \dfrac{R^3}{r^5}(2z^2 - x^2 - y^2 ) 
]
\]
and
\[
  \boldsymbol{H} =  [0, 0 , \frac{3}{ \frac{\mu_{\rm in}}{\mu_{\rm out}}   +2 }]
\]
inside the sphere, where $R$ is the radius of the sphere, $r = \sqrt{ x^2 + y^2 + z^2}$ \cite{PhysicsLibreTextsElectrostatics}, $\mu_{\rm in}$ and $\mu_{\rm out}$ are the permeability inside and outside the sphere respectively.
The permeability outside the sphere is set between $0.5 <\mu_{\rm out} < 1.5$, the permeability of the sphere $\mu_{\rm in} = 1 $, and the $z$ component of the uniform magnetic field is $H_z^0=1$. The permeability $\mu$ is approximated by $\mu = S(800 F)  (\mu_{\rm out} -1) + 1  $ and $\mu = S(2000 F)  (\mu_{\rm out} -1) + 1  .$ The input of the neural network is $[x,\mu_{\rm out},S(\beta F), \nabla S(50F) ]$  with $\beta = [200,300,400,500]$.

\begin{figure}[h]
\begin{subfigure}[t]{0.32\textwidth}
\centering
    \includegraphics[width=\textwidth,height=\textwidth]{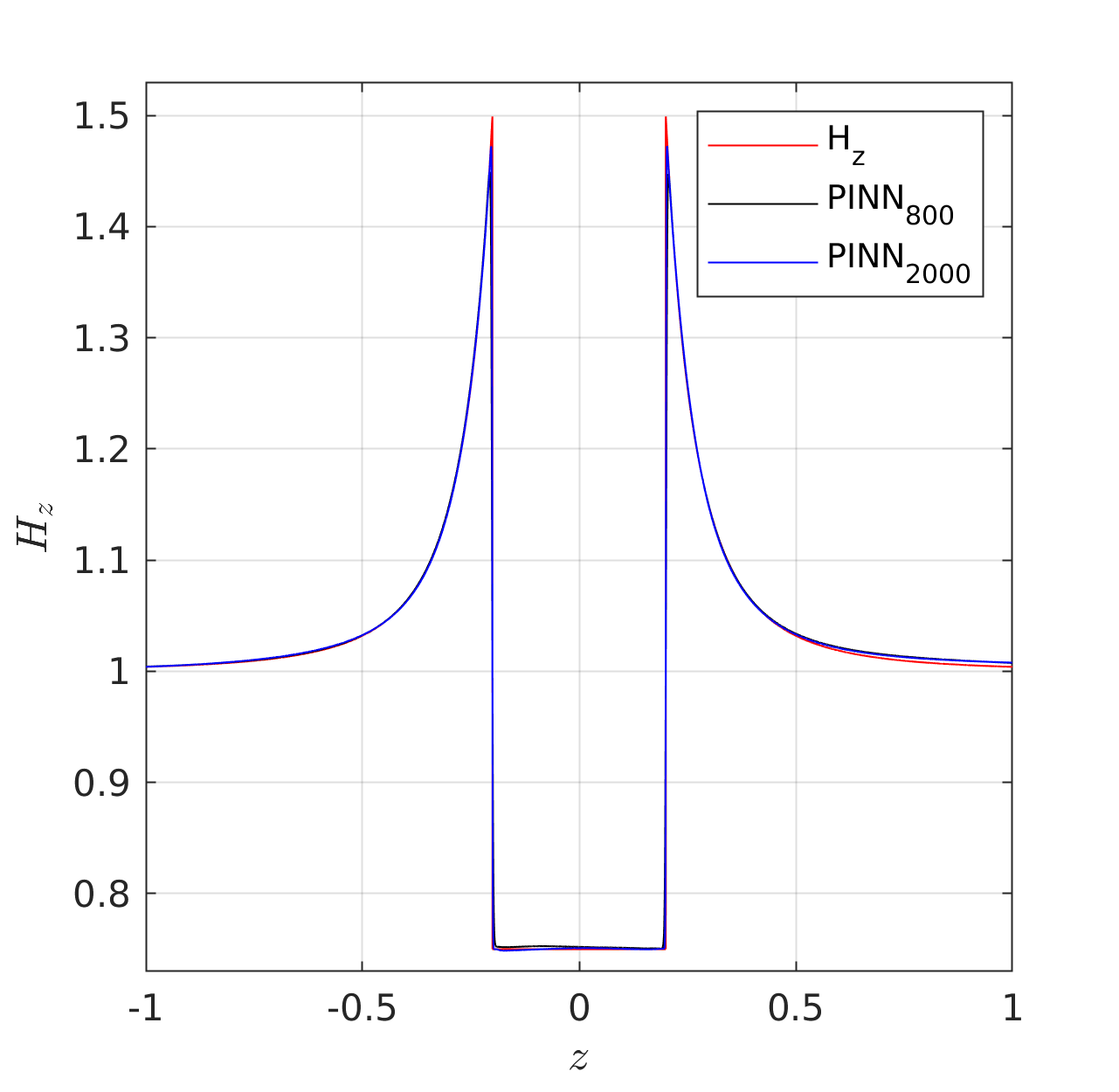}
\caption{}
\end{subfigure}
\begin{subfigure}[t]{0.32\textwidth}
    \includegraphics[width=\textwidth,height=\textwidth]{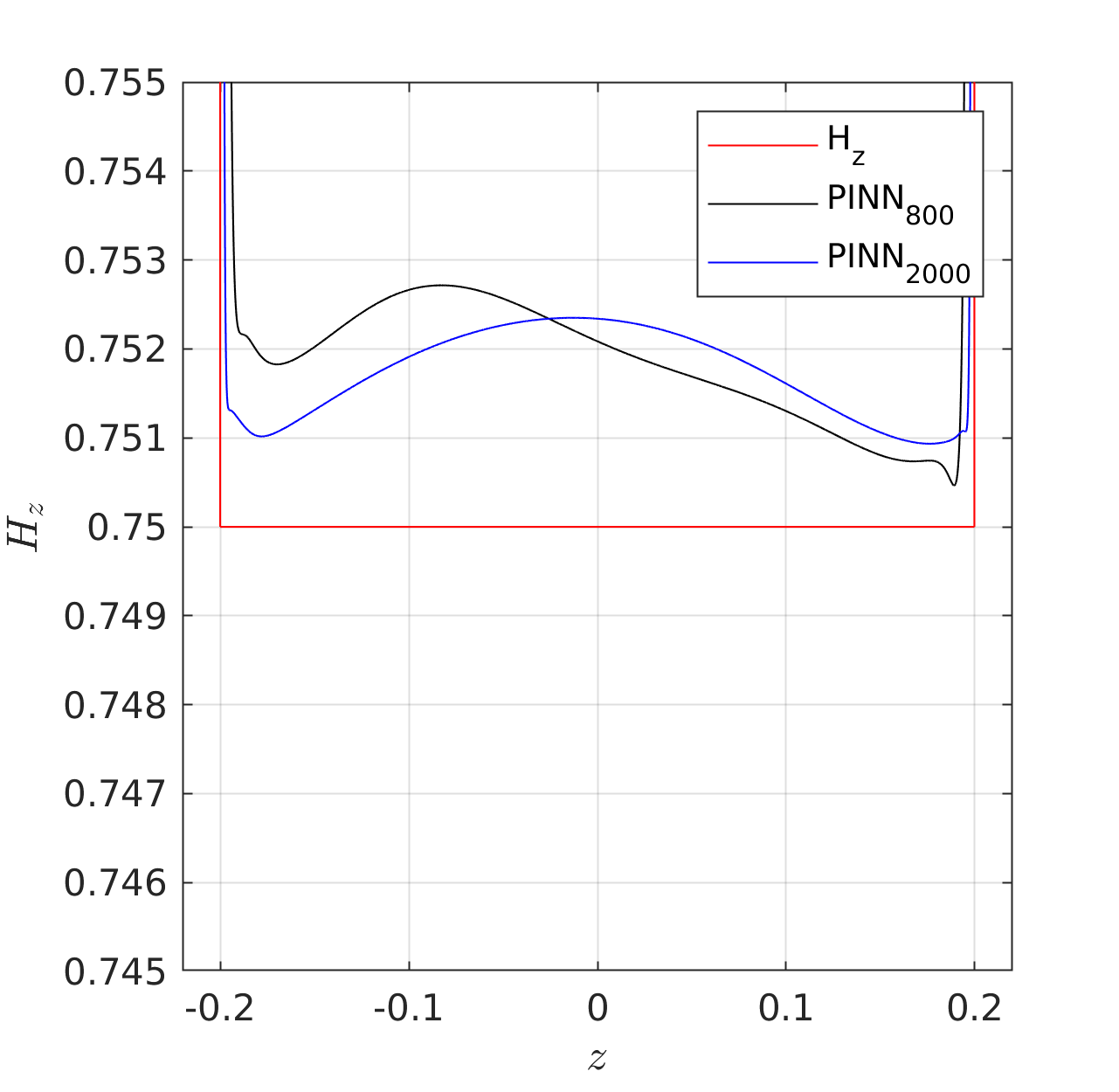}
\caption{}
\end{subfigure}
\begin{subfigure}[t]{0.32\textwidth}
    \includegraphics[width=\textwidth,height=\textwidth]{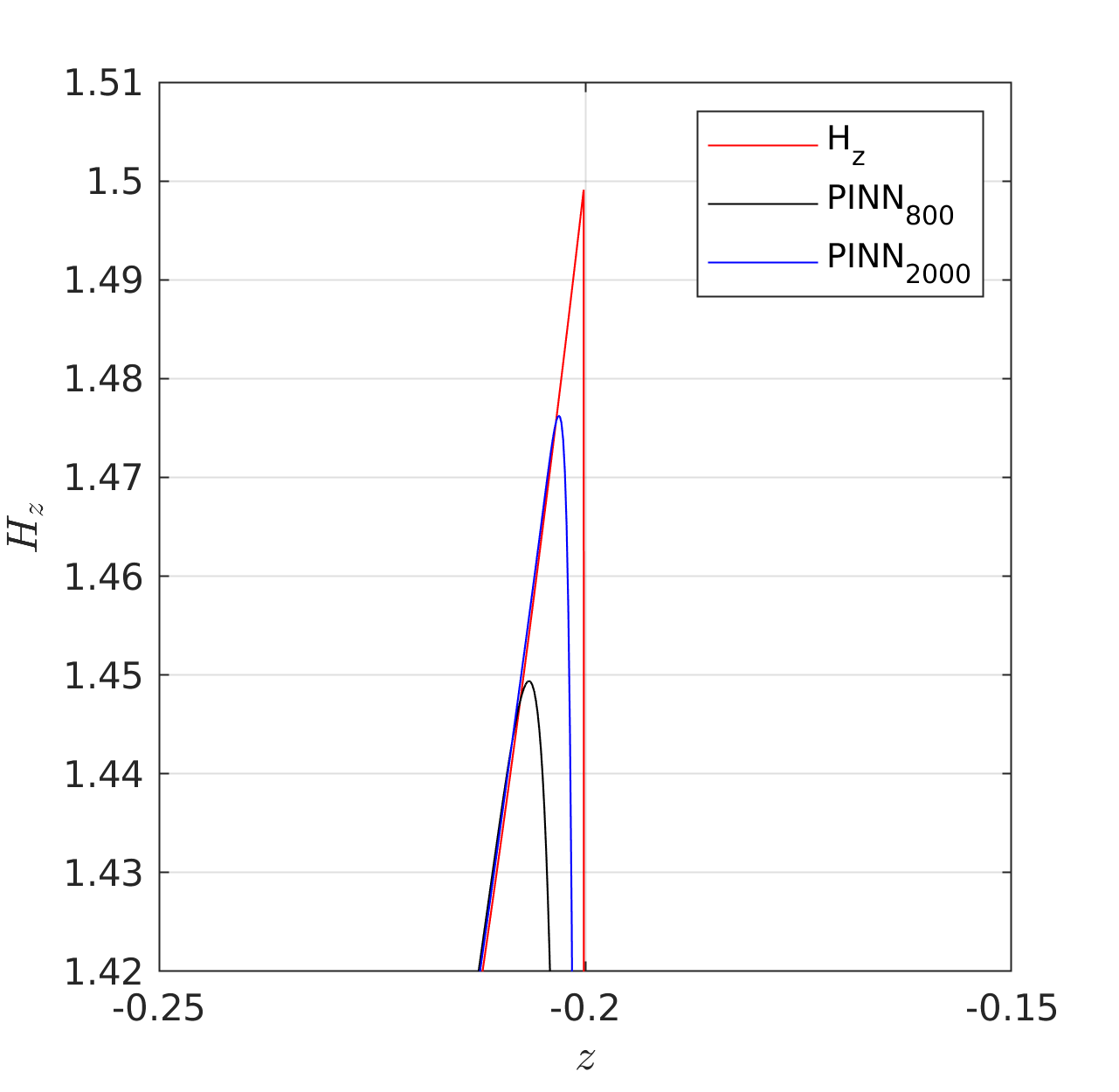}
\caption{}
\end{subfigure}
    \caption{Comparison of the exact solution $H_z$ with the approximated solution obtained by the proposed PINN for the parametric problem of a sphere inside a uniform magnetic field, at $x=y=0$, for $\mu_{\rm out}=0.5$, and an approximate Heaviside function $\hat{H}=S(\alpha F)$, with $a= 800$ and $2000$: (a) overall view of the entire field; (b) zoom-in around the sphere; (c) zoom-in around the interface. }
  \label{fig:exact_sphere1}
  \end{figure}
  
\begin{figure}[h]
\begin{subfigure}[t]{0.32\textwidth}
\centering
    \includegraphics[width=\textwidth,height=\textwidth]{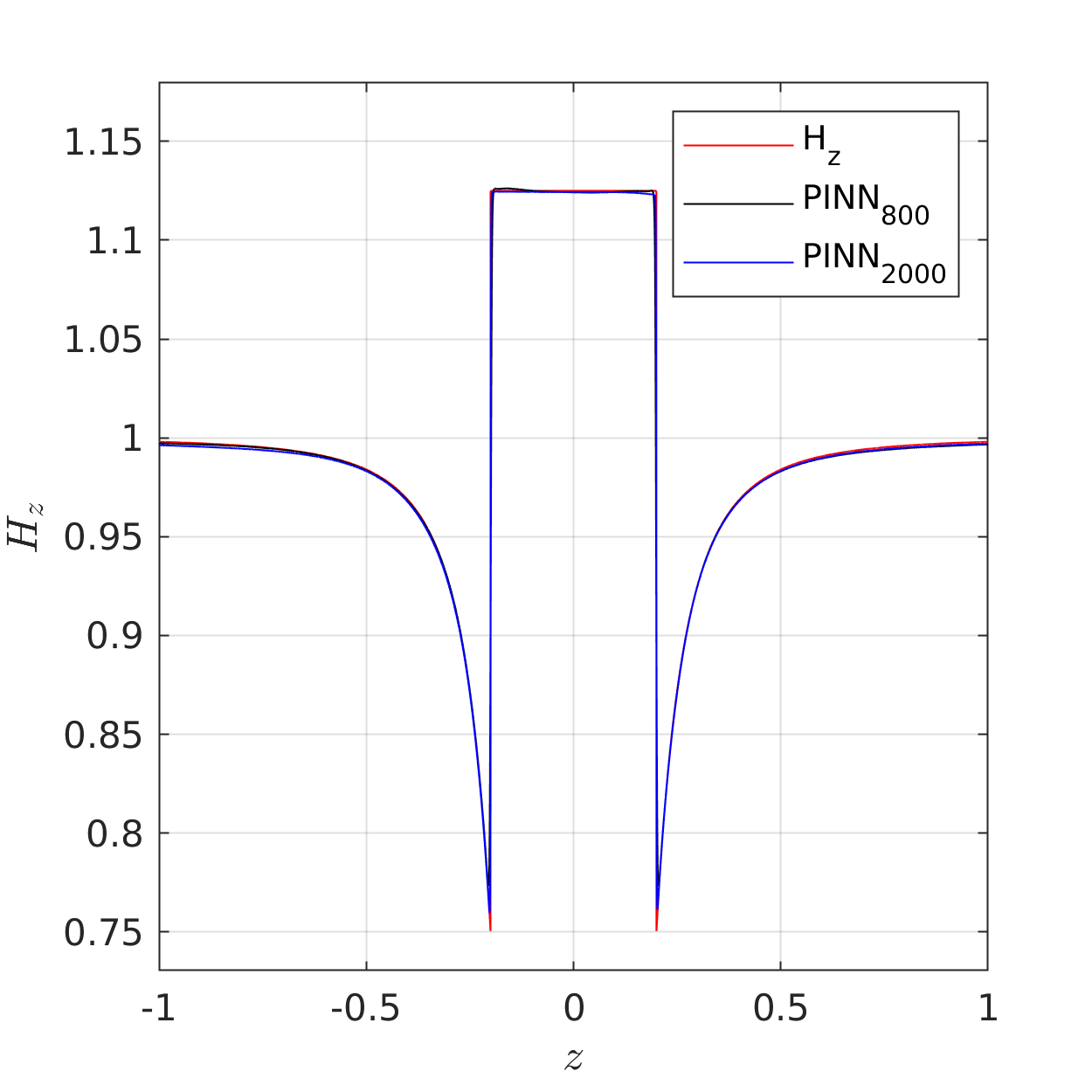}
\caption{}
\end{subfigure}
\begin{subfigure}[t]{0.32\textwidth}
    \includegraphics[width=\textwidth,height=\textwidth]{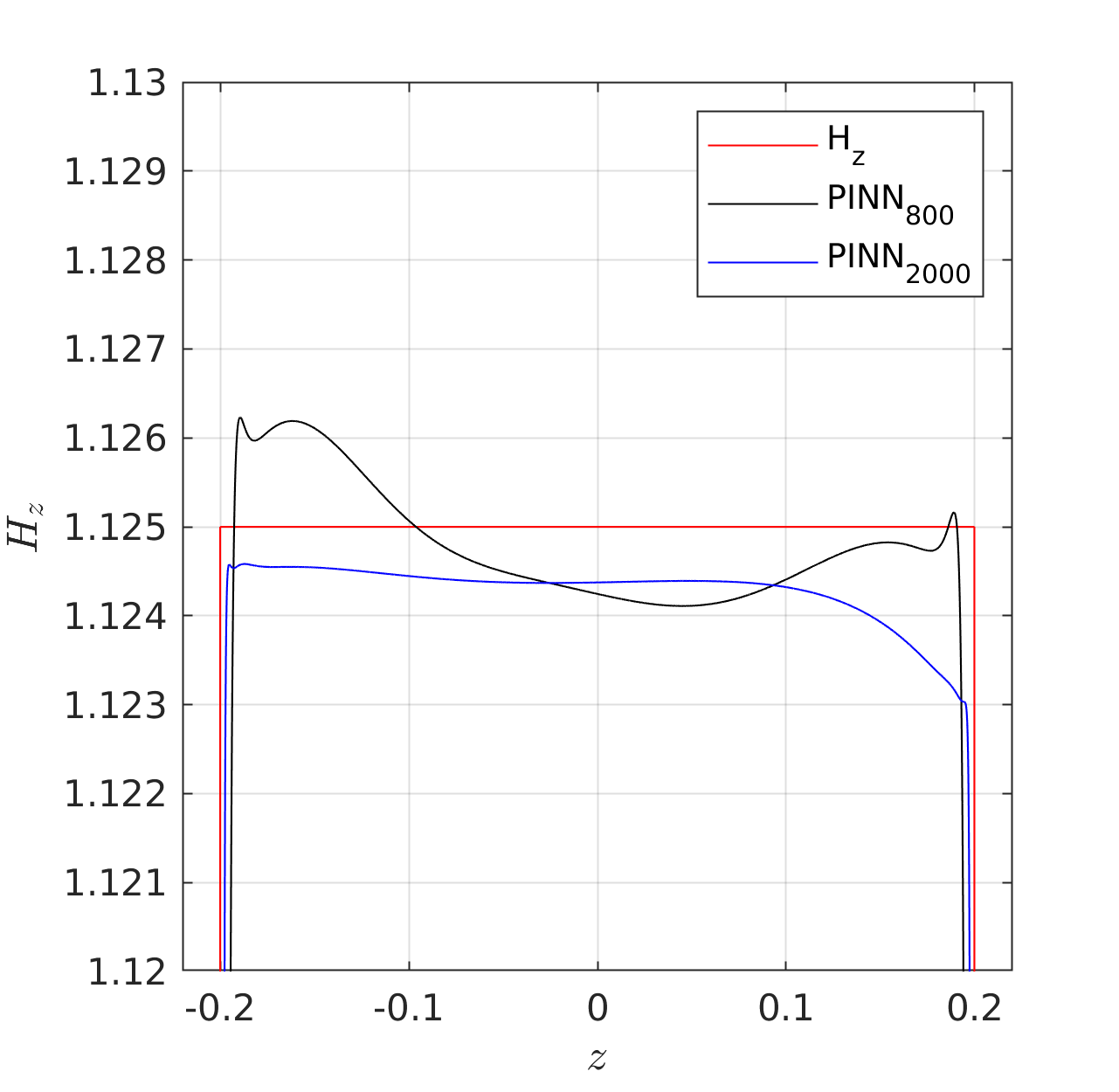}
\caption{}
\end{subfigure}
\begin{subfigure}[t]{0.32\textwidth}
    \includegraphics[width=\textwidth,height=\textwidth]{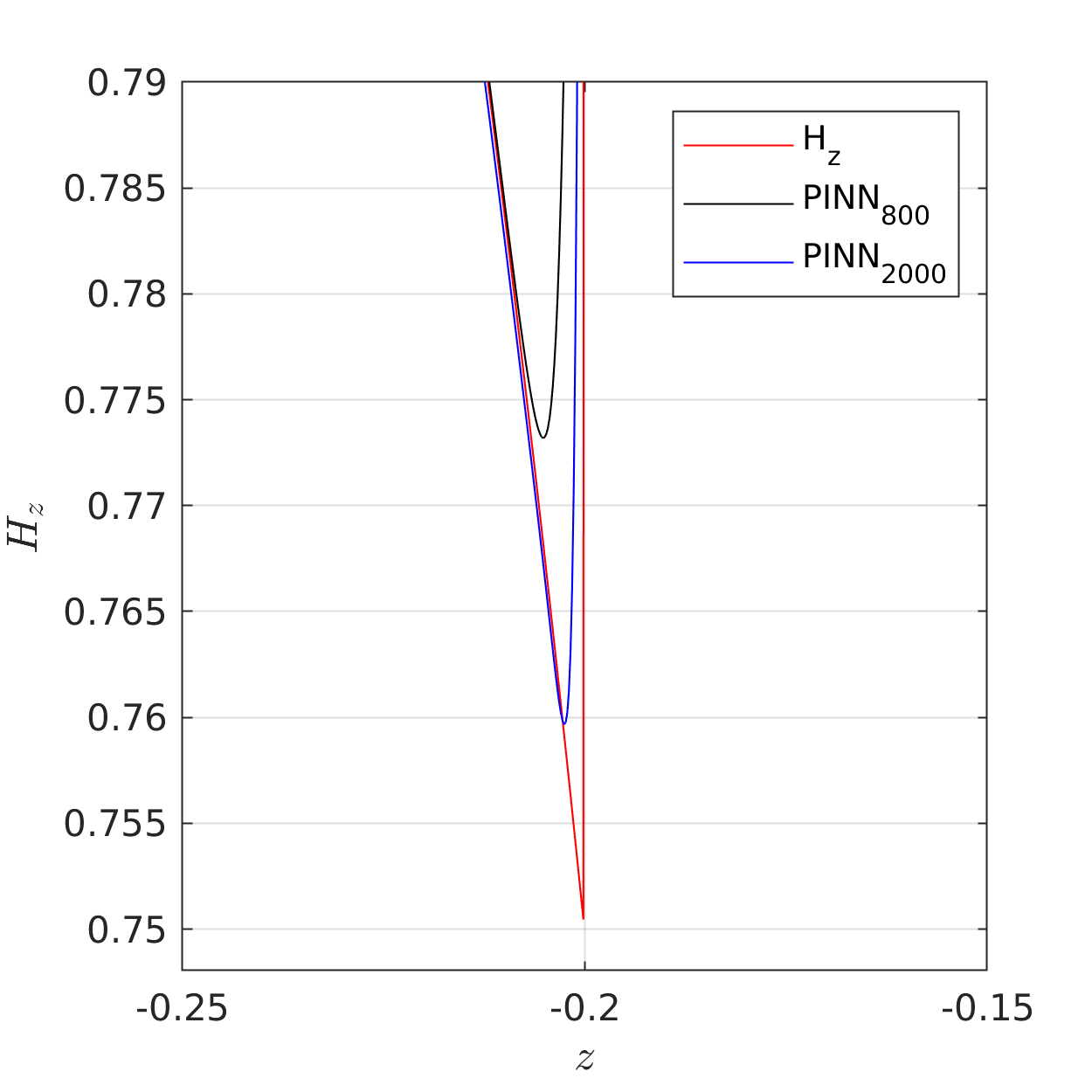}
\caption{}
\end{subfigure}
    \caption{Comparison of the exact solution $H_z$ and the approximated solution obtained using the proposed PINN for the parametric problem of a sphere inside a uniform magnetic field, at $x=y=0$, for $\mu_{out}=1.5$, and an approximate Heaviside function $\hat{H}=S(\alpha F)$, with $a= 800$ and $2000$: (a) overall view of the entire field; (b) zoom-in around the sphere; (c) zoom-in around the interface. }

    \label{fig:exact_sphere}
  \end{figure}
The results illustrated in figure \ref{fig:exact_sphere1} and~\ref{fig:exact_sphere}  show the validity of the Heaviside function approximation. As can be seen, the solution tends toward the exact solution on the interface for a steeper $\hat{H}$ approximation.
  
\subsection{Steady-state parametric problem of two concentric spheres inside a cube}

In this problem, we place two concentric spheres with respective radii $r=0.1$ and $r=0.4$ inside a unit cube. We impose the Dirichlet boundary condition on the magnetic field $\boldsymbol{H}=[0,0,1]$ on the four sides of the cube, and we impose the homogeneous Dirichlet boundary condition on the electric field $\boldsymbol{E} = \boldsymbol{0}$ on the bottom and top sides of the cube. The permeability of the inner sphere is $\mu_{\rm si}=1$, the permeability of the outer sphere ranges between $0.5<\mu_{\rm so}<1.5$, and the permeability outside the spheres is $\mu_{\rm out} = \mu_{\rm so}+1.$ 
The approximate Heaviside functions used are $\hat{H} = S(200 F_i(\boldsymbol{x}))$ and $\hat{H} = S(500 F_i(x))$, where $F_i$ is the level-set function to each of sphere surfaces. The input of the neural network is $\x_{\rm in} = [\x, \mu_{\rm so} , h(\beta) , \nabla h(50)] $, where $h(k)=S(k F_1(x)) + S(k F_2(x))$ and $\beta = [50,100,150,200]$.

\begin{figure}[H]
\begin{subfigure}[t]{0.5\textwidth}
\centering

    \includegraphics[width=\textwidth,height=\textwidth]{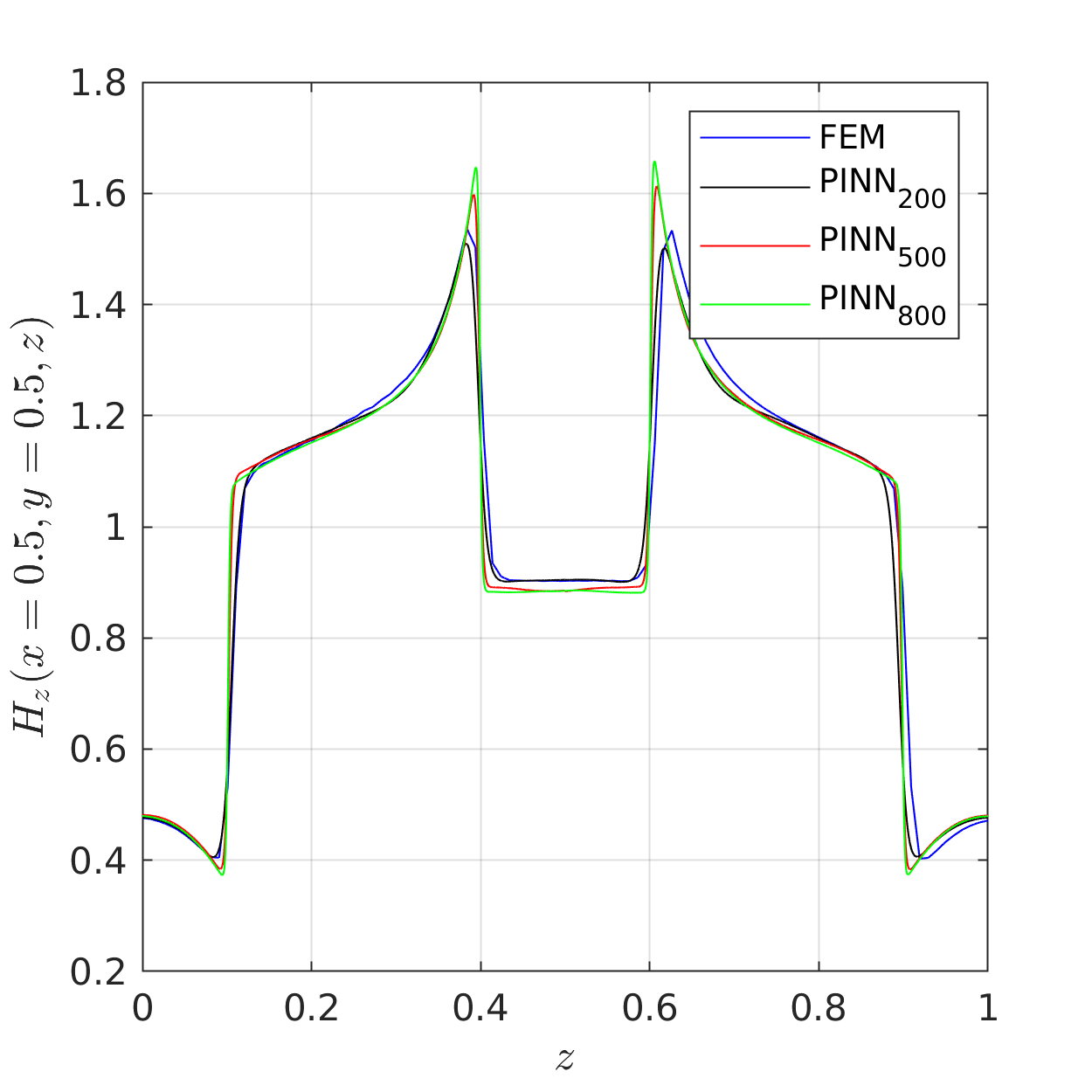}
\caption{}
\end{subfigure}
\begin{subfigure}[t]{0.5\textwidth}
\centering
    \includegraphics[width=\textwidth,height=\textwidth]{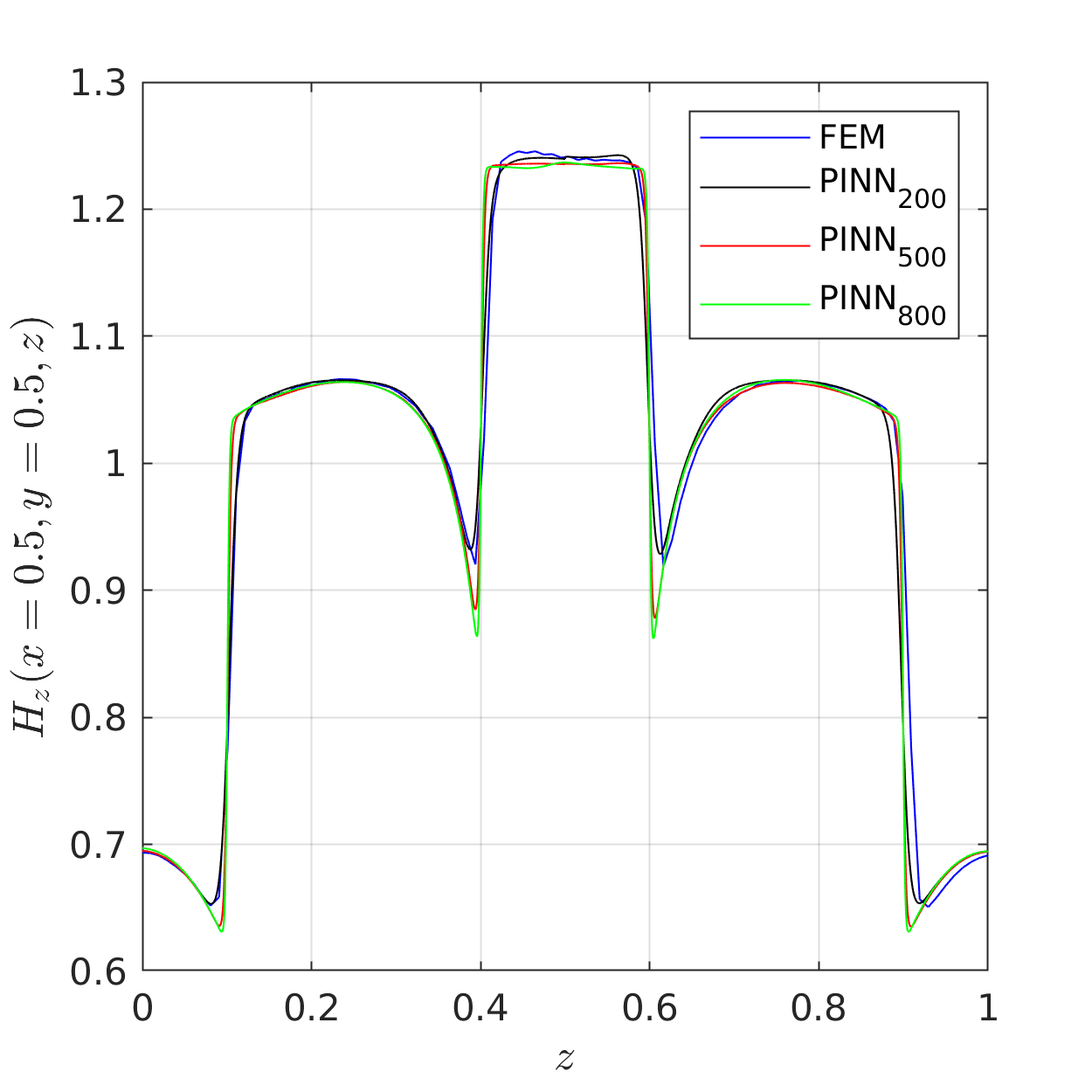}
\caption{}
\end{subfigure}

    \caption{$H_z$ at $x=y=0.5$ for the parametric steady-state problem of two concentric spheres inside a unit cube: (a) $\mu_{\rm so}=0.5$; (b) $\mu_{\rm so} = 1.5$. }
    \label{fig:22int_disc_mu}
  \end{figure}

  The results illustrated in figure~\ref{fig:22int_disc_mu} show the capability of the neural network to easily handle multiple material interfaces without any modification to the architecture.
 
\subsection{Steady-state parametric problem of an ellipsoid inside a cube}

\begin{figure}[ht!]
\begin{subfigure}[t]{0.32\textwidth}
\centering
    \includegraphics[width=\textwidth,height=\textwidth,valign=b]{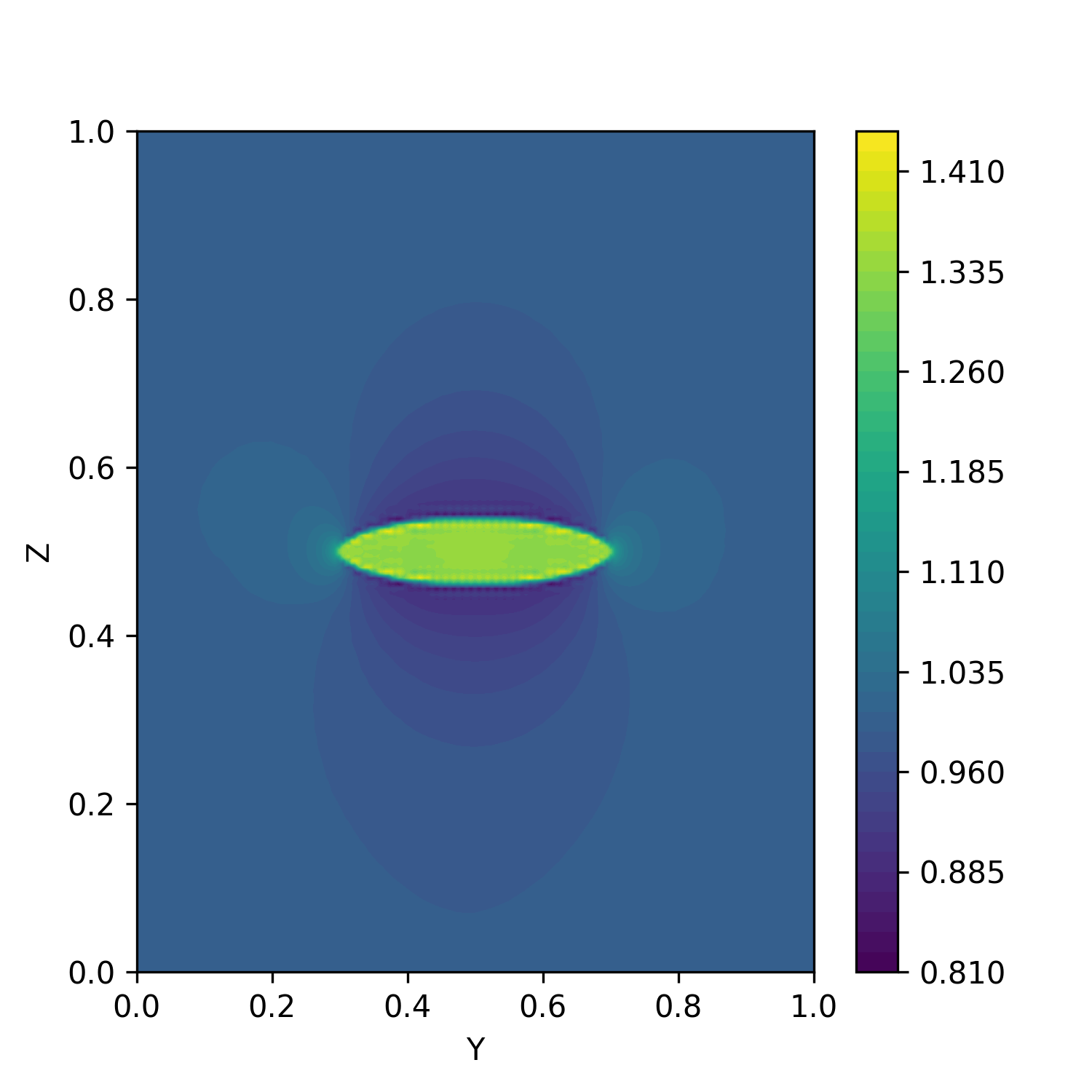}
\caption{}
\end{subfigure}
\begin{subfigure}[t]{0.32\textwidth}
\centering
    \includegraphics[width=\textwidth,height=\textwidth,valign=b]{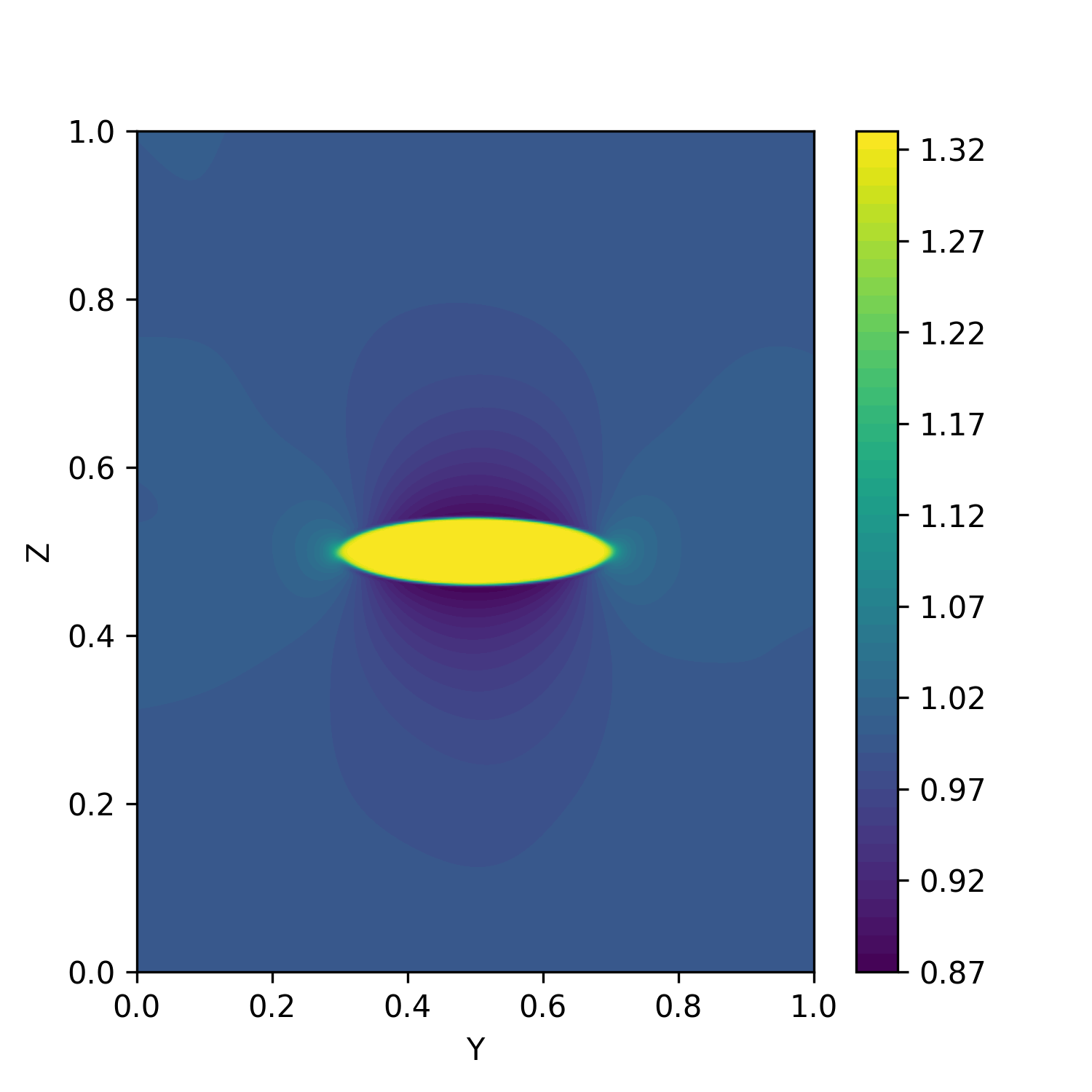}
\caption{}
\end{subfigure}
\begin{subfigure}[t]{0.32\textwidth}
\centering
    \includegraphics[width=0.9\textwidth,height=0.95\textwidth,valign=b]{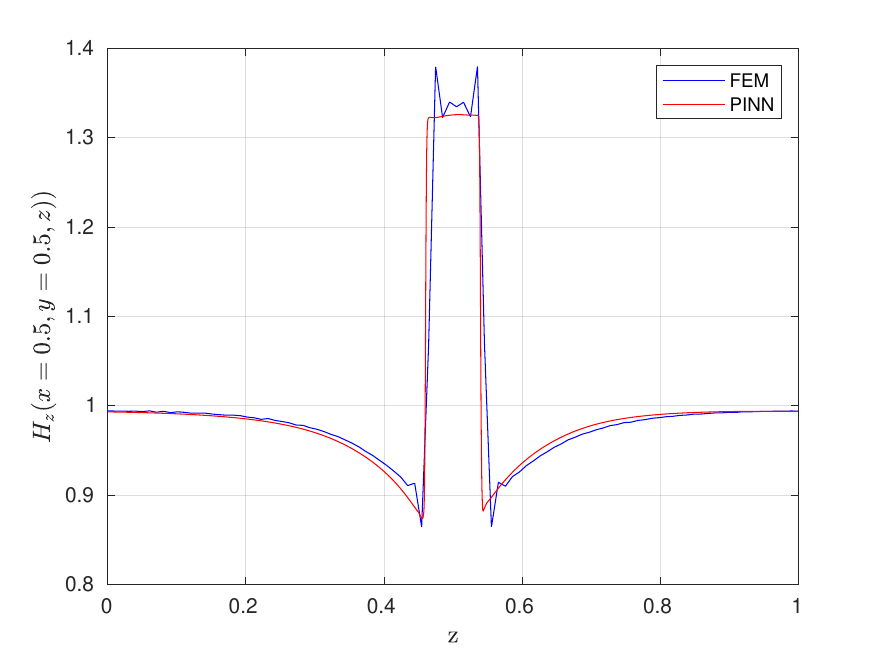}
\caption{}
\end{subfigure}
    \caption{The $z$-component of the magnetic field $H_z$ for the steady-state parametric ellipsoid problem, with $a=0.2$ at $x=0.5$: (a) obtained using the FEM method; (b) obtained using the proposed PINN; (c) The $z-$component of the magnetic field $H_z$ at $x=y=0.5$ obtained by the FEM and the PINN methods. }
    \label{fig:shape_02}
  \end{figure}

\begin{figure}[ht!]
\begin{subfigure}[t]{0.32\textwidth}
\centering
    \includegraphics[width=\textwidth,height=\textwidth,valign=b]{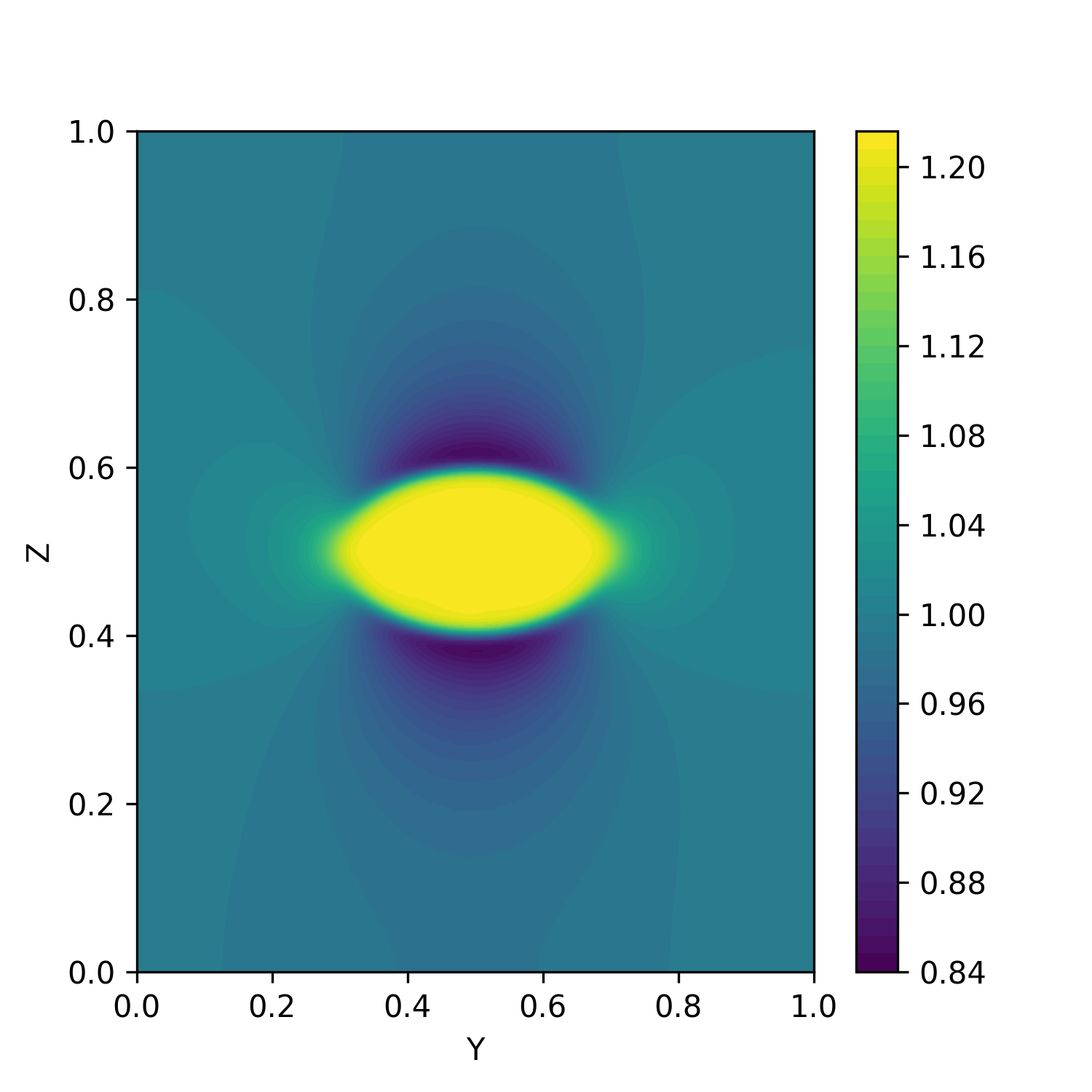}
\caption{}
\end{subfigure}
\begin{subfigure}[t]{0.32\textwidth}
\centering
    \includegraphics[width=\textwidth,height=\textwidth,valign=b]{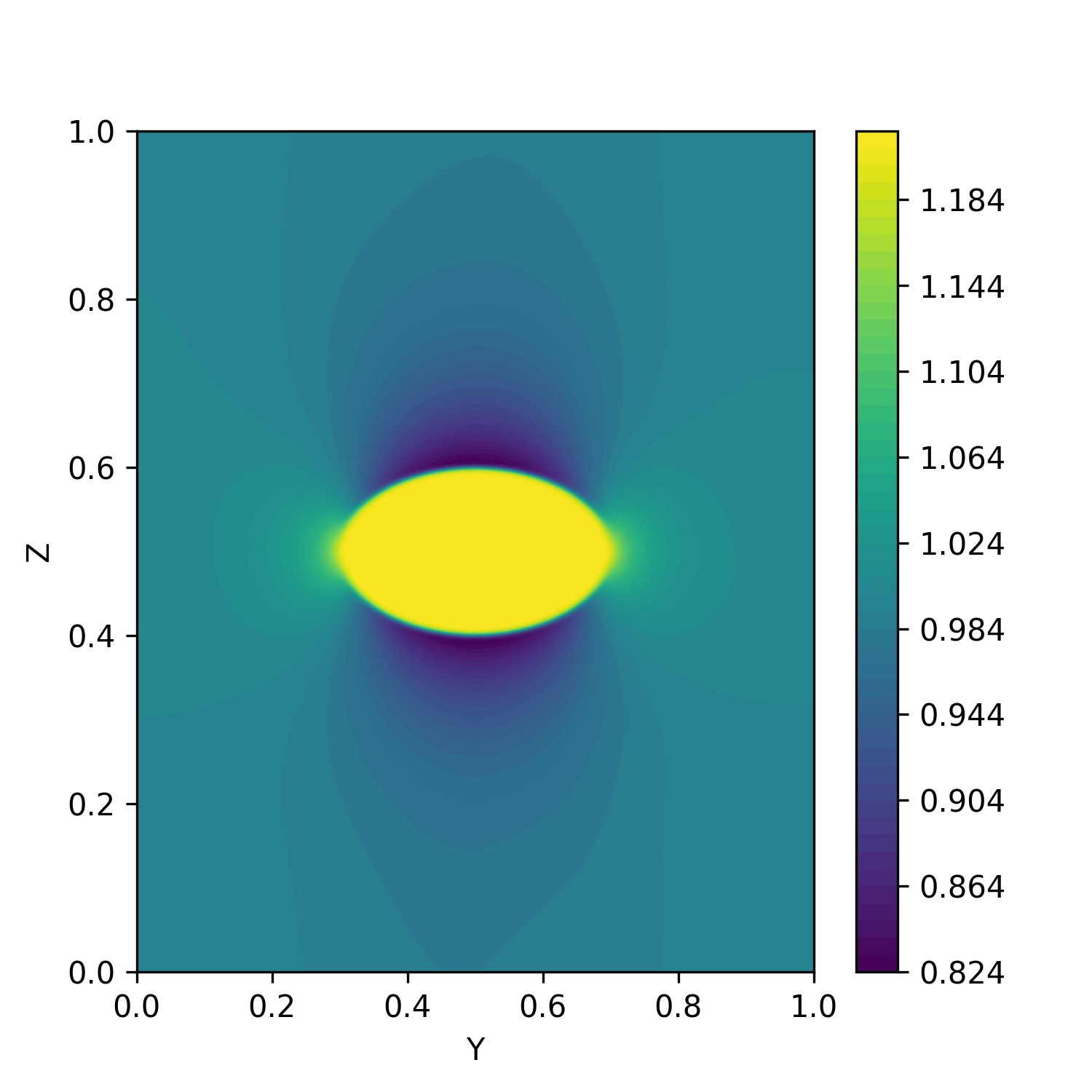}
\caption{}
\end{subfigure}
\begin{subfigure}[t]{0.32\textwidth}
\centering
    \includegraphics[width=0.9\textwidth,height=0.95\textwidth,valign=b]{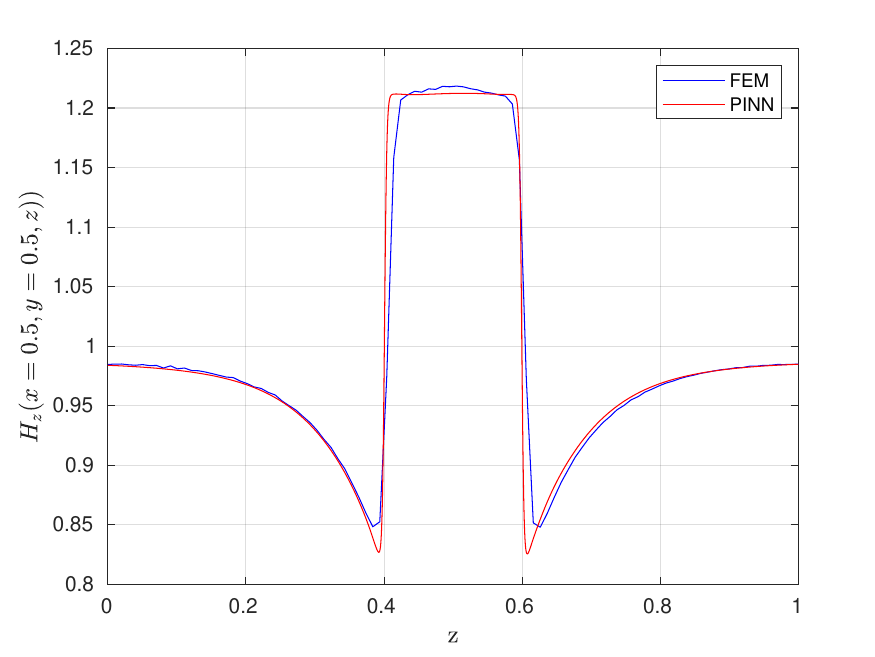}
\caption{}
\end{subfigure}
    \caption{The $z$-component of the magnetic field $H_z$ for the steady-state parametric ellipsoid problem, with $a=0.5$ at $x=0.5$: (a) obtained using the FEM method; (b) obtained using the proposed PINN; (c) The $z-$component of the magnetic field $H_z$ at $x=y=0.5$ obtained using the FEM and the PINN methods. }
    \label{fig:shape_05}
  \end{figure}

In this problem, we place an ellipsoid defined by $$F =\sqrt{ (x-0.5)^2 + (y-0.5)^2 + \dfrac{1}{a^2}(z-0.5)^2} - 0.2,$$ 
where $a$ is a parameter that varies between $ 0.1 < a < 1$, inside a unit cube.
The permeability outside the ellipsoid is $\mu = 1.5$, and $\mu = 1$ inside the ellipsoid.
The approximate Heaviside function used to represent the interface is $\hat{H} = S(500 F_i(\boldsymbol{x}))$ for the PINN method, and $\hat{H} = S(100 F_i(\boldsymbol{x}))$ for the FEM. The FEM method uses a smoother approximation to reduce oscillations at the interface. The results obtained using the two methods are illustrated in figures \ref{fig:shape_02} and \ref{fig:shape_05},
where we can see that the proposed architecture can capture the sharp gradients in varying positions, and it is more stable than the FEM in some cases, as seen in figure \ref{fig:shape_02}.
It is worth noting that the PINN proposed in \cite{nohra2} failed to converge for this problem.

\subsection{Transient parametric problem of a sphere inside a unit cube}
\begin{figure}[h]
\begin{subfigure}[t]{0.5\textwidth}
\centering
    \includegraphics[width=7cm,height=7cm]{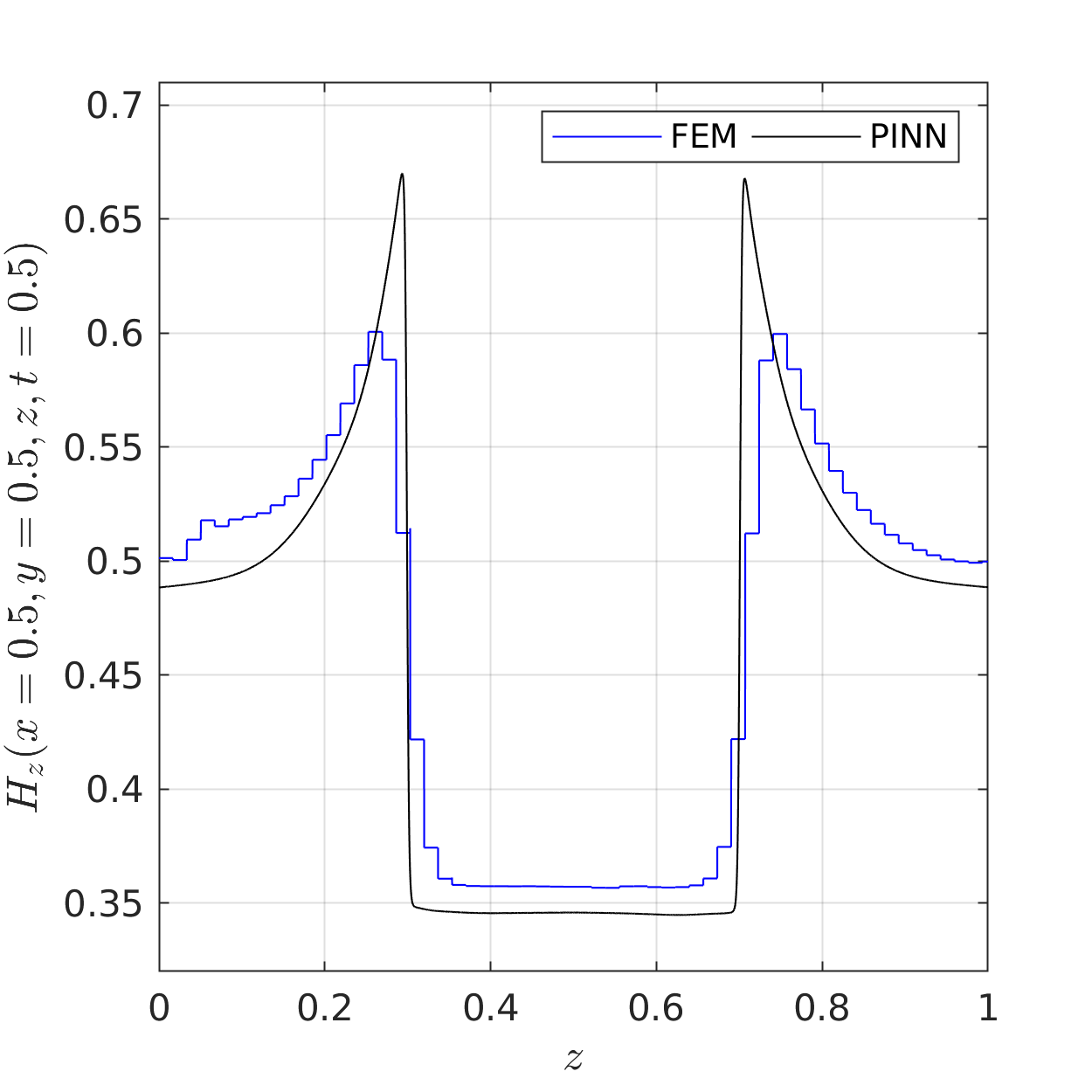}
\caption{}
\end{subfigure}
\begin{subfigure}[t]{0.5\textwidth}
\centering
    \includegraphics[width=7cm,height=7cm]{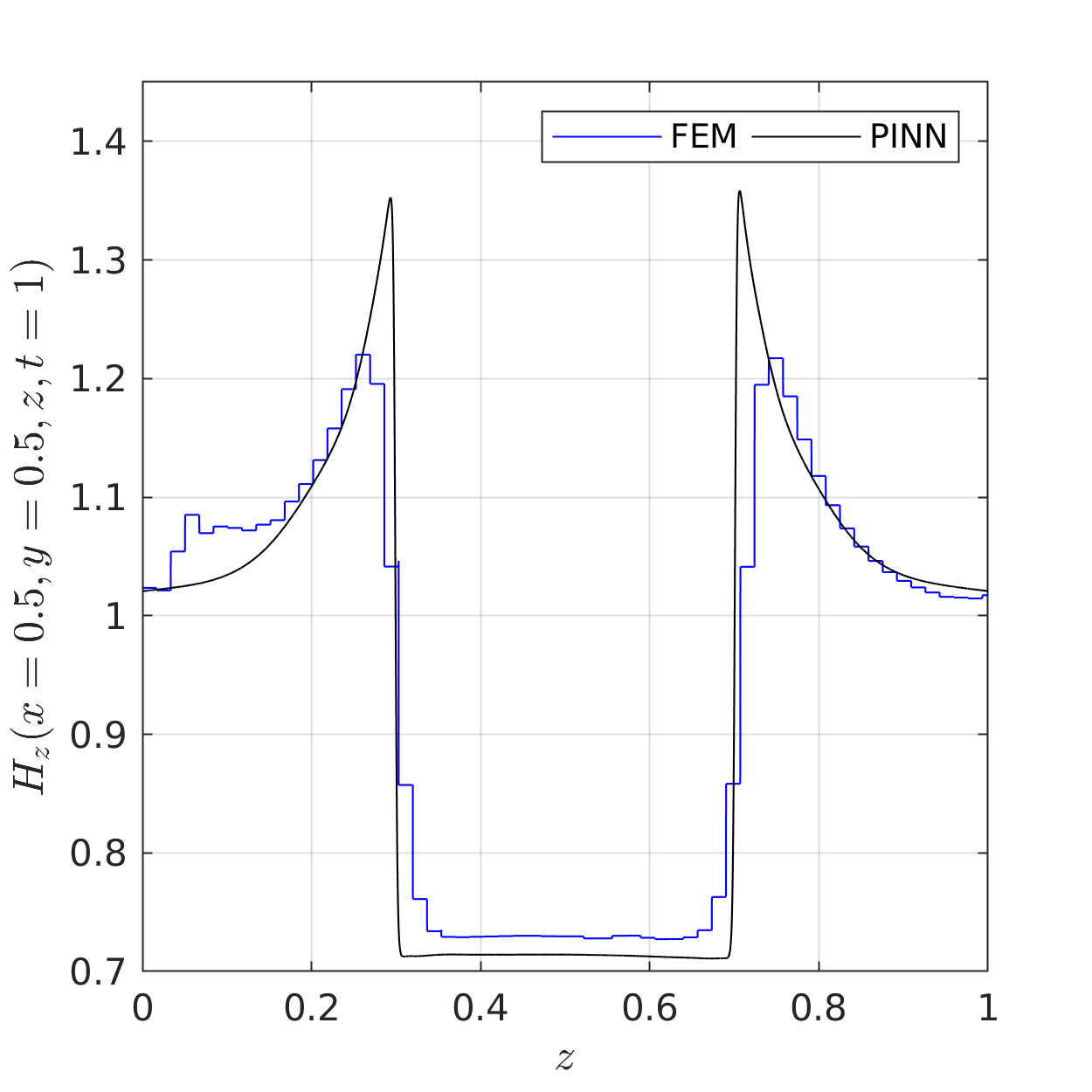}
\caption{}
\end{subfigure}
\begin{subfigure}[t]{0.5\textwidth}
\centering
    \includegraphics[width=7cm,height=7cm]{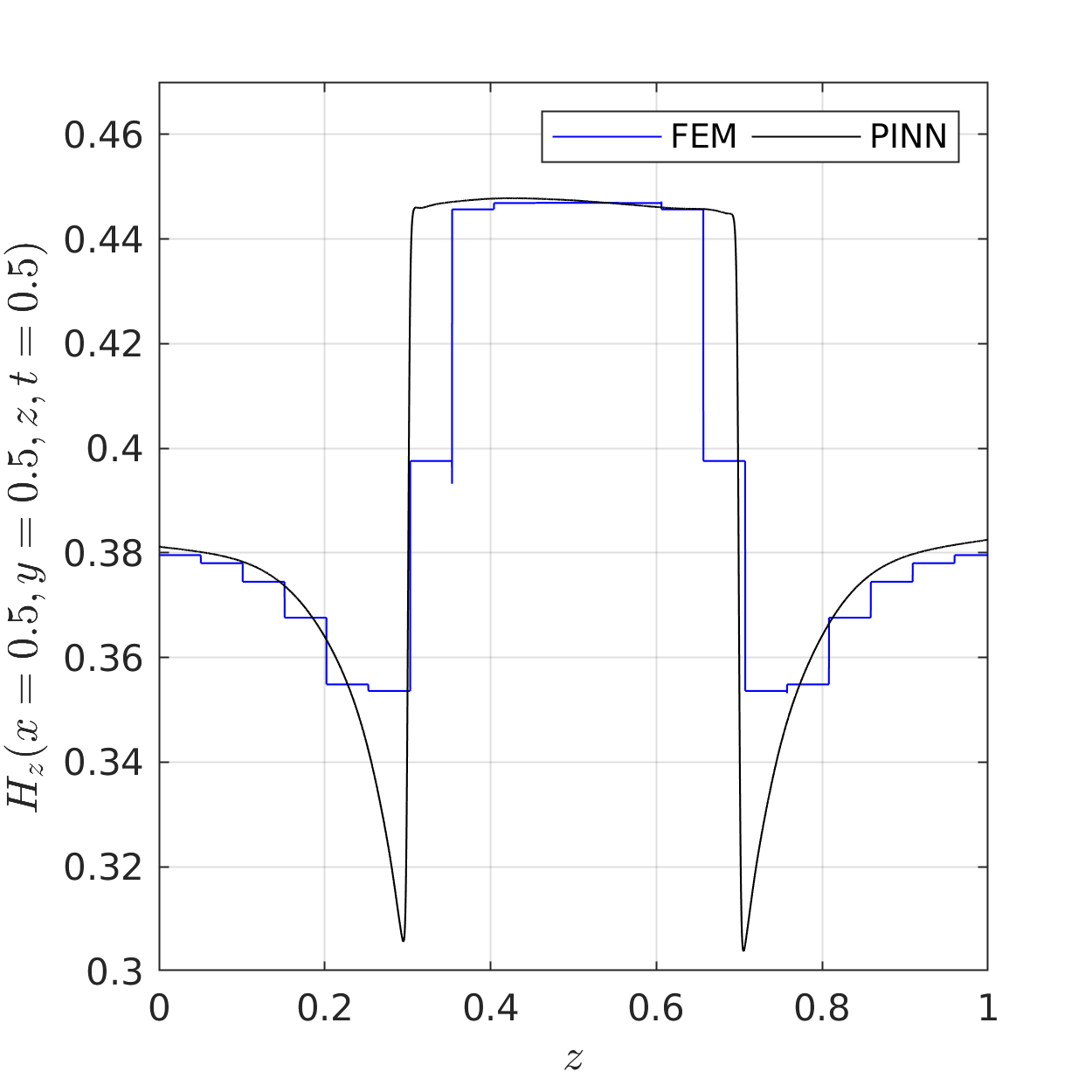}
\caption{}
\end{subfigure}
\begin{subfigure}[t]{0.5\textwidth}
\centering
    \includegraphics[width=7cm,height=7cm]{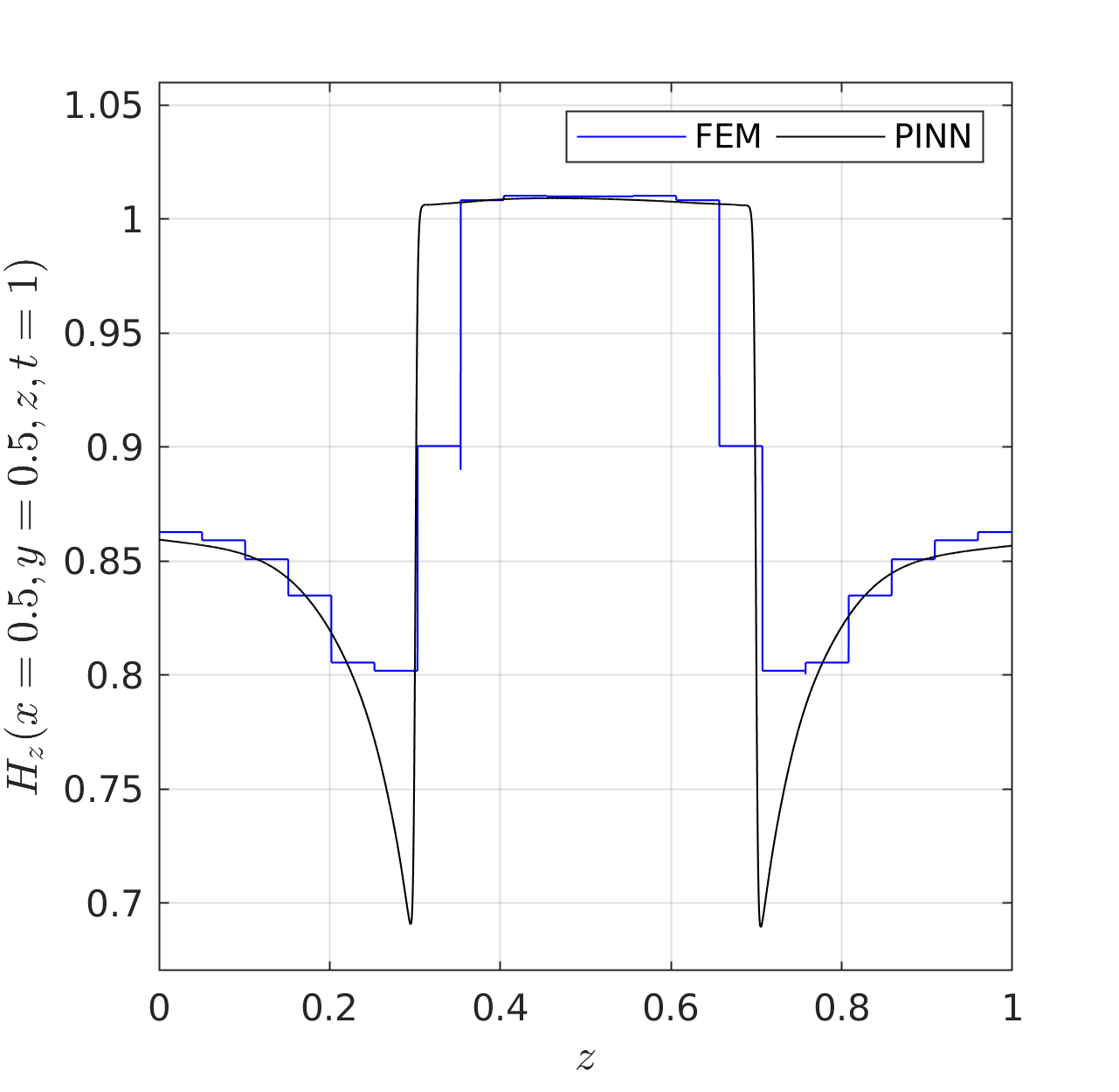}
\caption{}
\end{subfigure}
    \caption{$H_z$ at $x=y=0.5$ for the transient parametric problem of a sphere inside a unit cube obtained using the FEM and the proposed PINN: (a) $\mu_{\rm out}=0.5$ at time $t=0.5$; (b) $\mu_{\rm out}=0.5$ at time $t=1$; (c) $\mu_{\rm out}=1.5$ at time $t=0.5$; (c) $\mu_{\rm out}=1.5$ at time $t=1$. }
    \label{fig:time_disc_mu_2}
    \end{figure}

Finally, we validate the proposed PINN on the transient and parametric magnetic problem with a single interface. A sphere of radius $r=0.2$ and a permeability of $\mu=1$ is placed inside a unit cube with a permeability between $0.5<\mu_{\rm out}<1.5$. The resistivity is $\rho=1$ on the whole domain. The boundary conditions on the four sides of the cube are $\boldsymbol{H}=[0,0,t]$, and on the top and bottom sides of the cube are $\boldsymbol{E} = \boldsymbol{0}$.
We approximate the Heaviside function with $\hat{H} = S(800 F(\boldsymbol{x}))$. The results obtained using the proposed PINN are compared to the results obtained using the FEM in figure \ref{fig:time_disc_mu_2}. We notice good agreement between the results produced by the two methods, and we notice the presence of spurious solutions with the FEM, but not with the proposed PINN. The presence of spurious solutions is usually attributed to breaking the divergence-free property of the magnetic field. Since we can include this property in the loss function of the PINN, we obtain a more accurate solution than with the FEM.

\section{Conclusion}

We presented a numerical strategy based on PINNs for the parametric, static and transient electromagnetic problems in discontinuous media. The first-order formulation of Maxwell's equations was used based on the findings of~\cite{nohra2}, and the boundary and initial conditions were strongly imposed.
We used the level-set function to introduce high-frequencies near the interface, and in the normal direction of the interface, in the neural network inputs. This method was applied to numerous 3D parametric problems in static and dynamic regimes. 
The method was capable of handling sharp gradients at the interface, with varying interface locations, and with multiple interfaces, showing a promising alternative to traditional numerical methods like the FEM.



\clearpage
\bibliographystyle{abbrv} 
\bibliography{max}
\end{document}